\documentclass{jfm}

\usepackage{graphicx}
\usepackage{epstopdf, epsfig}
\usepackage[utf8]{inputenc}
\usepackage[T1]{fontenc}
\usepackage{lmodern}
\usepackage{natbib}
\usepackage{amsfonts,amssymb,amsmath}
\usepackage{here}
\usepackage{color}
\usepackage{soul}
\usepackage[normalem]{ulem}
\usepackage{cancel}
\usepackage{yfonts}
\usepackage{mathrsfs}
\usepackage{bm}
\usepackage{calligra}
\usepackage{physics}
 \usepackage[nocfg]{nomencl}
\makenomenclature

\usepackage[colorlinks=true]{hyperref}
\hypersetup{citecolor=blue,linkcolor=blue}

\newcommand{\x}{{\boldsymbol x}}
\newcommand{\xp}{\mathbf {x}_p}
\newcommand{\vv}{{\boldsymbol v}}
\newcommand{\uu}{{\boldsymbol u}}
\newcommand{\UU}{{\boldsymbol U}}
\newcommand{\up}{\mathbf {u}}
\newcommand{\g}{{\boldsymbol g}}

\newcommand{\vp}{\mathbf {v}_p}
\newcommand{\R}{\mathbb R}
\newcommand{\dotv}{\boldsymbol \cdot}
\newcommand{\nabv}{\boldsymbol \nabla}

\newcommand{\Uv}{\boldsymbol {\upsilon}_v}
\newcommand{\Ua}{\boldsymbol {\upsilon}_a}

\newcommand{\taut}{\boldsymbol \sigma}
\newcommand{\sigt}{\boldsymbol \tau}

\newcommand{\ratiov}{\eta}
\newcommand{\rhosw}{\rho_{sw}}
\newcommand{\rhow}{\rho_{w}}
\newcommand{\vchars}{v_{o}}
\newcommand{\vchar}{u_{o}}

\newcommand{\uchar}{u_{o}}
\newcommand{\Tchar}{T_{o}}
\newcommand{\xchar}{x_{o}}
\newcommand{\rchar}{r_{o}}
\newcommand{\mchar}{m_{o}}
\newcommand{\tchar}{t_{o}}
\newcommand{\fchar}{f_{o}}
\newcommand{\Rchar}{\mathcal{R}_{o}}
\newcommand{\Mchar}{\mathcal{M}_{o}}
\newcommand{\TTchar}{\mathcal{T}_{o}}

\newcommand{\Trate}{\mathcal{T}}

\newcommand{\taugrav}{\tau_{g}}
\newcommand{\tauR}{\tau_{R}}
\newcommand{\taum}{\tau_{m}}
\newcommand{\tauT}{\tau_{T}}
\newcommand{\tauI}{\tau_{D}}

\newcommand{\tauIchar}{\tau_{Do}}

\newcommand{\tauIs}{\tau_D'}
\newcommand{\vs}{\vv'}
\newcommand{\Ts}{T'}
\newcommand{\xs}{\x'}
\newcommand{\rs}{r'}
\newcommand{\ms}{m'}
\newcommand{\ts}{t'}
\newcommand{\fs}{f'}

\newcommand{\gs}{\g'}
\newcommand{\Rs}{\mathcal{R}'}
\newcommand{\Ms}{\mathcal{M}'}
\newcommand{\TTs}{\mathcal{T}'}

\newcommand{\agravs}{\alpha_g}
\newcommand{\agrav}{\frac{1}{{\rm Fr}^2}}
\newcommand{\aI}{\alpha_D}
\newcommand{\aR}{\alpha_R}
\newcommand{\am}{\alpha_m}
\newcommand{\aT}{\alpha_T}

\newcommand{\rhochar}{\rho_{o}}

\newcommand{\pchar}{p_{o}}
\newcommand{\muchar}{\mu_{o}}
\newcommand{\kappachar}{\kappa_{o}}
\newcommand{\cchar}{c_{o}}
\newcommand{\Tas}{T_a'}
\newcommand{\rhos}{\rho'}
\newcommand{\qs}{q'}
\newcommand{\us}{\uu'}
\newcommand{\sigts}{\sigts'}
\newcommand{\ps}{p_a'}
\newcommand{\mus}{\mu'}
\newcommand{\kappas}{\kappa'}
\newcommand{\Ma}{{\rm Ma}}
\newcommand{\Reyn}{{\rm Re}}
\newcommand{\Pe}{{\rm Pe_T}}
\newcommand{\Pem}{{\rm Pe}_v}

\newcommand{\prodm}{\pi_m}
\newcommand{\produ}{\boldsymbol{\pi}_{\uu}}

\newcommand{\prode}{\pi_e}

\newcommand{\MS}{{\mathbb M}_S}
\newcommand{\ti}{\tilde{t}}
\newcommand{\HS}{{\mathbb H}_S}
\newcommand{\US}{{\mathbb U}_S}
\newcommand{\TSL}{{\mathbb T}_{S,L}}
\newcommand{\TSS}{{\mathbb T}_{S,S}}
\newcommand{\XX}{X}
\newcommand{\VV}{V}
\newcommand{\msw}{m_{sw}}

\shorttitle{An Eulerian model for sea spray}
\shortauthor{F. Veron \& L. Mieussens}

\title{An Eulerian model for sea spray transport and evaporation}

\author{
Fabrice Veron\aff{1}
 \and Luc Mieussens\aff{2}}

\affiliation{
\aff{1} School of Marine Science and Policy, University of Delaware, Newark DE, USA\\
\aff{2} Bordeaux INP, Univ. Bordeaux, CNRS, IMB, UMR 5251, F-33400 Talence,France 
}

\begin{document}

\maketitle

\begin{abstract}
Reliable estimates of the fluxes of momentum, heat, and moisture at the air-sea interface are essential for accurate long term climate projections, as well as the prediction of short term weather events such as tropical cyclones. In recent years, it has been suggested that these estimates need to incorporate an accurate description of the transport of sea spray within the atmospheric boundary layer and the  drop-induced fluxes of moisture, momentum, and heat, so that the resulting effects on the atmospheric flow can be evaluated. In this paper we propose a model based on a theoretical and mathematical framework inspired from kinetic gas theory. This approach reconciles the Lagrangian nature of the spray transport with the Eulerian description of the atmosphere. In turn, this enables a relatively straightforward inclusion of the spray fluxes and the resulting spray effects on the atmospheric flow. A comprehensive dimensional analysis has led us to identify the spray effects that are most likely to influence the temperature, moisture, and speed of the airflow. We also provide an example application to illustrate the capabilities of the model in specific environmental conditions. Finally, suggestions for future work are offered.

\end{abstract}


\section{Introduction}
The fluxes of sensible heat, water vapour, and momentum at the air-sea interface are crucial ingredients in the overall energy and mass balances of both the oceans and the atmosphere. As such, these air-sea fluxes are important boundary conditions for atmospheric and oceanic models that attempt to capture the physics and evolution of weather and climate. These fluxes can be relatively well estimated in moderate wind speed conditions using so-called bulk parametrizations, which are extensions of the classical turbulent boundary layer flow theory applied to the air-sea interface \citep[e.g.][]{Fairall:1996c}. However, the presence of surface waves and, in particular, breaking events produce 
phenomena such as bubble injections and turbulence generation on the water side, and airflow separation and spray generation on the air side. In turn, these phenomena cause the flows to depart from the classical boundary layer descriptions \citep[e.g.][]{Thorpe:1993, Melville:1994, Anis:1995, Melville:1996, Terray:1996, Veron:2001, Hristov:1998, Hare:1997, Adrian:1991, Edson:1998, Donelan:2004}. Consequently, in conditions when surface wave breaking is pervasive, be it because of intense winds or the presence of coastlines, air-sea fluxes are not fully resolved. In fact, in the past decade, measurements at high wind speeds confirm that simple air-sea flux parametrizations are still lacking for conditions in which breaking is ubiquitous \citep{Drennan:2007, Zhang:2008, Bell:2012}.

The knowledge gap in high wind conditions can be in part attributed to our incomplete understanding of the role of sea spray. Indeed, in strongly forced conditions, it is well known that significant amounts of sea spray are generated at the surface of the ocean and get subsequently transported into the marine atmospheric boundary layer. In turn, these spray drops are believed to affect the multiple air-sea fluxes. However, there are still large uncertainties as to the amount of heat, moisture, and momentum that the drops exchange with the atmosphere while suspended \citep{Andreas:1990, Veron:2015}.  In this paper, we develop a model which accounts for the direct effects of the spray on the atmosphere.

Sea spray is composed of sea water drops ejected from the sea surface, generally due to breaking waves and breaking related phenomena such as bubble entrainment and whitecaps.  Spray is mainly formed through two pathways. The first, occurs when bubbles entrained by breaking waves rise to the surface and burst \citep[see][]{Lewis:2004, DeLeeuw:2011}. When the bubble surface shatters, small film drops with radii less than $O(1)\; \mu m$ are formed \citep{Spiel:1998, Fuentes:2010a}. The subsequent collapse of the cavity that is leftover leads to the ejection of so-called jet drops with radii of order $O(1-40)\; \mu m$ \citep{Wu:2002, Walls:2015, Wang:2017}. The second process occurs when the wind is sufficiently strong to tear off water from the top of waves or from splashing from plunging breaking waves. Early measurements by \citet{Koga:1981} showed that globules or filaments, produced on the front face of the waves, breakup under the shearing effect of the wind and generate a distribution of large drops \citep{Marmottant:2004, Mueller:2009c}. These are referred to as spume drops \citep{Andreas:1995, Veron:2015}. This filament breakup process is likely accompanied by a bag breakup phenomena whereby small water canopies are inflated by the wind and shatter \citep{Veron:2012, Troitskaya:2018a}. The splash production of drops from plunging breaking waves has only been recently investigated \citep{Erinin:2019}. In general, spume and splash drops have radii larger than $O(20)\; \mu m$.

Once ejected from the ocean surface, these drops are transported and dispersed in the atmospheric boundary layer where they interact with the turbulent airflow and exchange momentum, heat, and moisture with the atmosphere \citep[e.g.][]{Fairall:1994, Edson:1996, Andreas:1999, Andreas:2002, Andreas:2008, Mueller:2014b}. It is through these exchanges that sea spray is believed to play a critical role in the overall air-sea heat and moisture fluxes which fuels the intensity of tropical cyclones \citep{Bao:2000, Andreas:2001, Andreas:2004, Andreas:2008, Andreas:2010, Liu:2010, Bianco:2011, Rosenfeld:2012}. 

In order to quantify the spray effects on the air-sea fluxes of heat and moisture, both an accurate quantification of the source of spray, and a suitable knowledge of the thermodynamics and transport of aqueous drops in air, are required. With regards to the thermodynamics, seminal works by \citet{Pruppacher:1978} and \citet{ Andreas:1990} have guided numerous studies, \citep[e.g.][]{Rouault:1991, Andreas:1992, Edson:1994, Andreas:1995, Edson:1996, Mestayer:1996,Makin:1998,Andreas:2001,VanEijk:2001, Lewis:2004, Barenblatt:2005, Mueller:2010a, Mueller:2010b, Bao:2011, Mueller:2014a, Mueller:2014b}. Likewise, in order to assess the momentum exchanges between drops and the airflow, a comprehensive knowledge of the drop dynamics at ejection and during its flight through the turbulent atmosphere is needed. Nowadays, drop's thermodynamics and their (Lagrangian) dynamics are quite well understood and modeled. Thus, a complete assessment of the spray-mediated air-sea fluxes relies on the proper estimate of the amount of spray present in, and ejected into, the airflow. This is typically quantified using the so-called 
 the sea spray generation function, or spray source function, i.e. the number distribution of spray drops generated at the ocean surface (per unit surface area and unit time). However, this source function has proven difficult to directly measure, especially 
 for the larger drops which may reside within a wave height of the surface and thus cannot be easily detected by typical fixed height measurements. Consequently, despite recent laboratory measurements \citep{Veron:2012, Ortiz:2016, Troitskaya:2018a}, theoretical estimates and parametrizations of the spray generation function are not well constrained, particularly for field applications and large drop radii. Yet, these large and heavy drops may have the largest potential for exchanging momentum with the wind \citep{Andreas:1998, Fairall:2009}. Small film drops with radii smaller than $O(1)\; \mu m$ eventually partake in the global marine aerosol flux but are expected to have a minimal influence on the overall heat and momentum air-sea fluxes \citep{Andreas:1990}. Accordingly, in this paper, we consider spray drops with radii in the range $O(1)\; \mu m$ to $O(1)\; mm$.

 Overall, the currently incomplete quantification of the spray source function (or equivalently the spray concentration and velocity distributions) is a substantial limiting factor in properly assessing the impact of sea spray on the total air-sea fluxes of heat, moisture, and momentum.  In this paper however, we do not focus on the details of the spray generation process, or the spray generation function and its shortcomings. The generation function serves as a boundary condition for the model proposed therein, and the choice of which generation function is best suited for a particular set of conditions is left to the user. 
 
 Still, based on spray concentration functions commonly used \citep[e.g.][]{Fairall:1994, Andreas:2002, Andreas:2010} it is possible to estimate, even if roughly, a number of useful parameters. For example, one can estimate that even in wind speed conditions reaching approximately $50\; m s^{-1}$, the total spray volume fraction (i.e. the volume occupied by the liquid drops in a unit  volume of air) is on the order of $O(10^{-5})$ \footnote{Note that the  the spray generation function of \citet{Fairall:1994} or \citet{Andreas:2002} have not been validated at such high wind speeds; however, they are generally regarded as reliable and typically serve as a reference. As such, the order of magnitude estimate presented here still provides informative overview.}. In these conditions, drop collisions and coalescence are expected to be extremely improbable. However, because mass and heat capacity of water drops are larger than that of the air, estimates of the spray mass and heat content ratios show that they can reach up to $O(1)\%$ and $O(10)\%$ respectively. This simply means that while the spray drops occupy a negligible volume, they carry a significant amount of the total mass and heat contained in the air-spray mixture. Thus, one should expect that the spray could indeed play a substantial role in the total air-sea momentum, heat, and moisture fluxes. 
 
  As hinted above, an additional major challenge associated with modeling spray-mediated exchanges is the Lagrangian nature of transport of the drops through the atmosphere, whereas the spray concentration function and the airflow are traditionally described and modeled in an Eulerian frame of reference \citep{Edson:1994, Edson:1996, Mueller:2009b, Mueller:2010b, Shpund:2011, Shpund:2012, Richter:2013, RichterPoF:2013, HelgansIJMF:2016, RichterPoF:2014, Peng:2017, Peng:2019}. 
 
 In this paper, we present a theoretical framework, originally developed in the context of kinetic gas theory, with which we reconcile the Eulerian and Lagrangian approaches. This kinetic (``discrete mechanics'') technique is also, in part, motivated and legitimized by the spray volume concentration estimate above which indicates that in most conditions, the spray drop concentration is so-called ``thin''. We believe that the model presented here will prove useful in simulating the transport of spray drops and the exchanges of heat, momentum, and moisture between the drops and the atmosphere.

\section{Sea spray evolution equations}

To be able to describe the influence of the sea spray on the atmosphere, both the motion and the thermodynamics of single drops needs to be examined. This is inherently a \textit{Lagrangian} approach where we assume that each spray drop is made of a single fluid with density $\rho_p$, and is spherical with radius $r_p$ and mass $m_p=\rho_p\frac{4}{3}\pi r_p^3$. Each drop is then uniquely defined by its position $\xp \in \R^3$; its velocity $\vp \in \R^3$, its radius  $r_p \in [0, +\infty[$, its mass  $m_p \in [0, +\infty[$, and its temperature $T_p \in [0, +\infty[$.  We note that the drop mass and radius are independent variables because sea spray drops are made of sea water but evaporation  only transfers pure water (water vapour) between the spray drop and the air. Consequently, at a particular time, given only the drop radius but ignoring its time history, one cannot state precisely the drop mass. For instance, a small drop could be newly formed and have the density of sea water $\rho_{sw}$, or it could be a large drop that has substantially evaporated, and thus would be saltier with a density superior to that of sea water.

\subsection{Motion of individual spray drops}
\label{subsec:motion}
The motion of individual spray drops is given by:
\begin{align}\label{eq:Stokes_flow1}
\frac{d \xp(t)}{dt }&=\vp(t) \\ 
 m_p(t) \frac{d  \vp(t) }{dt }  &= 
\boldsymbol {\mathcal{F}}(t).
\label{eq:Stokes_flow2} 
\end{align}
The first equation relates the location of the drop $\xp(t)$ to
its velocity $\vp(t)$. The second equation is Newton's Second law
where $\boldsymbol {\mathcal{F}}$ accounts for gravity and air frictional effects
\begin{equation}
\boldsymbol {\mathcal{F}}=m_p \g+{\boldsymbol F}
\label{eq:drag_law}
\end{equation}
but includes neither higher order effects such as the Basset history and the added mass terms, nor Faxen, Saffman, and Magnus effects. These equations follow individual drops and are therefore \textit{Lagrangian} in nature.  Gravity is noted $\g$ and $\rho$ is the density of the (moist) air. The drag force exerted by the fluid on the drop is described with a general drag law
\begin{equation}
{\boldsymbol F}=\frac{1}{2} \rho C_{D} A \left( \up-\vp\right)  \left| \mathbf {v}_s \right|
\label{eq:drag_law2}
\end{equation}
where  $\up$ is the velocity of the air (at the location of the drop as if it were absent), $C_{D}$ is the drag coefficient of the drop and $A=\pi r_p^2$ its cross sectional area. This drag coefficient generally depends on the drop Reynolds number  $Re_{p}= \frac{2r_p\left| \mathbf {v}_s \right|}{\nu}
$, with $\mathbf {v}_s=\vp-\up$, usually referred to as the slip velocity, and $\nu=\mu/\rho$  the kinematic viscosity of air. It is convenient to express the drag coefficient as $C_{D}=\frac{24}{Re_{p}}\Xi$, where $\Xi$ is an empirical correction factor which accounts for departures from the Stokes flow as $Re_{p}$ increases and the flow around drops becomes turbulent \citep[e.g.][]{Clift:1970}.

Equivalently, one can express the viscous drag force in a more compact form
\begin{equation}
{\boldsymbol F}=\frac{m_p}{\tauI}  \left(\up-\vp \right)  
\label{eq:drag_law2a}
\end{equation}
where $\tauI$ is the drop
response time to the drag force. In this case, just like the drag coefficient, $\tauI$ can be expressed as $\tauI=\frac{\tau_{p}}{\Xi}$ where $\tau_{p}=\frac{2r_p^2\rho_p}{9\mu}$ is the drop Stokes time scale. Naturally, in Stokes flow with $Re_{p} \ll 1$, $\Xi\to 1$, $\tauI\to\tau_{p}$, $C_{D}\to\frac{24}{Re_{p}}$ and the viscous drag  ${\boldsymbol F}\to\frac{m_p}{\tau_{p}}\left(\up-\vp \right)= 6 \pi r_p \mu  \left(\up-\vp \right)$. Non-continuum effects can also be accounted for in the correction factor $\Xi$ through the use of a Cunningham slip coefficient \citep{Davies:1945}, but these non-continuum effects do not have a significant influence on the drag of drop radii larger than approximately 0.3-0.5 $\mu m$.
 
In the remainder of this paper, we distinguish between the following  derivatives:
$\frac{d}{dt}$  the derivative of a function of time only (Lagrangian for example), and $\frac{{ \mathrm{d}}}{{\mathrm{d} t}}= \frac{{ \partial   }}{{\partial t}} + \vp \dotv \nabv _{\x}$ the material derivative advected at the velocity of the drop (which is different from the traditional material derivative, $\frac{D}{D t}=\frac{ \partial}{\partial t}+\uu \dotv \nabv _{\x}$, where the advection is done by the Eulerian fluid velocity; see \citet{Maxey:1983} for example).  
Likewise, vector quantities will be noted in upright font when taken as Lagrangian variables  and noted in italics when representing Eulerian variables (e.g. $\up$ denotes the air velocity at the location of the drops and $\uu$ is the Eulerian air velocity as in Navier-Stokes equation below). The gradient operator $\nabv_{\x}$ is noted with a subscript $\x$ to indicate that the differentiation is with respect to physical space.

\subsection{Thermodynamics of individual spray drops}
In addition to the drop position and momentum, it is clear from equation~(\ref{eq:Stokes_flow2})
 that we need to determine the mass of the drop and thus its thermodynamical state. As sea spray drops evaporate, 
their mass change according to  (\citet{Pruppacher:1978}, their equation 13.9):
\begin{equation}
\frac{d m_p(t)}{dt}= 4 \pi r_p  D_v^* \left(\rho_v-\rho_{v,p}\right),
\end{equation}
where $\rho_v$ is the water vapour  density, and $D_v^*$ the  molecular diffusivity for water vapour modified to include non-continuum and high Reynolds number effects (see below).
In this model, the mass transfer results from the difference between the ambient vapour density and that directly adjacent to the surface of the drop, noted $\rho_{v,p}$. For the latter, it is generally necessary to account for the effects of drop curvature (surface tension), and salinity. With these effects explicitly written out, the evolution equation for the mass of a saline sea spray drop takes the following form (\citet{Pruppacher:1978} equation 13.26):
\begin{eqnarray}
\frac{d m_p(t)}{dt} &=&  \frac{4 \pi r_p  D_v^* M_w p_v^{\textrm{sat}} \left( T_a\right) }{ R T_a }  \nonumber \\
&\times& \left[ Q_{RH} - \frac{T_a}{T_p} \exp\left(\frac{L_v M_w }{R}\left(\frac{1 }{T_a}-\frac{1 }{T_p}\right)+\frac{2M_w \Gamma_p }{ R   \rho_w r_p T_p} - {\frac{I \Phi _s m_s \left( {M_w} / {M_s } \right)}{m_p - m_s }}\right) \right] \nonumber \\
 &=& \mathcal{M}(t)
\label{eq:DropRadius} 
\end{eqnarray}
where it is noted that the rate of variation  $\mathcal{M}$  depends on both the
fields of the carrier fluid and on the Lagrangian state of the
drop, i.e. $\mathcal{M}=\mathcal{M}(\rho,\up, T_a; \vp, r_p,m_p, T_p)$. In the equation above, $Q_{RH}$ is the fractional relative humidity; $R$ is the universal gas constant; $M_w$ and $M_s$ are the molecular weight of water and salt, respectively; $p_v^{\textrm{sat}}\left( T_a\right)$ is the saturation water vapour pressure at the air temperature $T_a$; $T_p$ is drop temperature assumed uniform inside the drop; $L_v$ is the latent heat of vaporization;  $\Gamma_p$ is the surface tension for a flat surface at the interface of the drop; $m_s$ is the mass of salt in the drop; $I$ is the number of ions salt dissociates into ($I$=2), and $\Phi_s$ is the osmotic coefficient.  In equation (\ref{eq:DropRadius}), the term between brackets [ ] simply represents the difference between the ambient humidity $Q_{RH}$ surrounding the drop and the humidity at the surface of the drop which results from three distinct effects that are reflected in the three terms inside the $exp$. The first term accounts for the drop temperature which differs from the ambient air temperature. The second term reflects curvature and surface tension effects. The last term accounts for salinity effects in the spray drop.

Noting that the mass of salt inside a spray drop remain constant (i.e. evaporation or condensation transfers only pure water), at a given time, 
\begin{equation}
m_p(t)=\frac{4}{3} \pi \rho_{sw} r_{p0}^3+\frac{4}{3} \pi \rho_{w} \left( {r_{p}(t)}^3 -r_{p0}^3 \right),
\end{equation}
where $\rho_{sw}$ is the density of sea water, $\rho_{w}$ is the density of (pure) water, and $r_{p0}$ is the radius of a drop at formation ($t=0$). In the equation above, the first term is the mass of the drop at formation and the second term represents the mass lost due to the evaporation of pure water. Incidentally, this computation shows that the mass of salt in the
drop (required in~(\ref{eq:DropRadius})) is $m_s = m_p(t) - \frac{4}{3} \pi r_p(t)^3\rho_{w}$. Thus, assuming that $\rho_w$ does not change substantially in time (because of temperature changes for example),
\begin{eqnarray}
\frac{d  m_p(t) }{dt } &=&4\pi r_{p}^2 \rho_{w} \frac{d r_p(t)}{dt}  \nonumber \\
&=& 4\pi r_{p}^2 \rho_{w} \mathcal{R}(t)  \nonumber \\
&=& \mathcal{M}(t),  \label{eq-M}
\end{eqnarray} 
which readily gives the evolution equation for the drop radius. 

Similarly, the temporal evolution of the drop temperature $T_p$ is given in \citet{Pruppacher:1978} (their equation 13.65):
\begin{equation}\label{eq:DropTempPruppacher}
m_p c_{ps} \frac{d T_p(t)}{dt} = 4\pi r_p   k_a^* \left(T_a-T_p\right) + L_v \frac{d  m_p(t) }{dt }
\end{equation}
which yields
\begin{eqnarray}\label{eq:DropTemp}
\frac{d T_p(t)}{dt} &=& \frac{4\pi r_p}{m_p c_{ps}} \left(  k_a^* \left(T_a-T_p\right) + L_v  D_v^* \left(\rho_v-\rho_{v,p}\right) \right) \nonumber \\
&=&  \mathcal{T}(t) 
\end{eqnarray}
 where the rate of variation  $\mathcal{T}=\mathcal{T}
 (\rho, \up, T_a; \vp, r_p, m_p, T_p)$  also depends on the 
 macroscopic fields of the carrier fluid and on the Lagrangian
 variables of the drop. In the expression for $\mathcal{T}$, we neglect high order effects related to temporal and spatial gradients in
the air temperature  \citep{Michaelides:1994}.
In the equation above, $c_{ps}$ is the specific heat of salty water, and $k_a^*$ the modified thermal conductivity of air. The two terms in $\mathcal{T}(t)$  represent the changes to the
drop temperature from the sensible and latent heat flux exchanges.

The modified diffusivity for water vapour, $D_v^*$, and the modified thermal conductivity of air, $k_a^*$, include noncontinuum effects and are respectively:
\begin{eqnarray}
D_v^{*}&=&\frac{f_vD_v} {  \frac{r_p}{r_p+\Delta _v} + \frac{D_v}{r_p \alpha _c}\left(\frac{2 \pi M_w}{R T_a} \right)^{1/2} }, \\ 
k_a^{*}&=&\frac{f_h k_a} {  \frac{r_p}{r_p+\Delta _T} + \frac{k_a}{r_p \alpha _T \rho _{a} c_{\textrm{p},a}}\left(\frac{2 \pi M_a}{R T_a} \right)^{1/2} }, 
\end{eqnarray}
where $D_v$ and $k_a$ are the molecular diffusivity for water
vapour and thermal conductivity of air, respectively.  Also, $M_a$ is the molecular weight of dry air; $c_{\textrm{p},a}$  and $\rho _{a}$ are respectively the specific heat (at constant pressure) and density of dry air, with constants $\Delta _v=8\times 10^{-8}$,  $\alpha_c=0.06$, $\Delta_T=2.16 \times 10^{-7}$, and $\alpha_T=0.7$. 
We also include ventilation coefficients $f_v=f_h=1+ \sqrt{Re_p} / 4$, which are corrections to the heat and water vapour diffusivity for large instantaneous drop Reynolds numbers, i.e. $Re_p >>1$ \citep{Edson:1994}.  These are analogous to the mean Sherwood and Nusselt numbers corrected using Ranz-Marshall correlations \citep{Ranz:1952}.

In short, in this Lagrangian approach, the system that governs the evolution of sea spray drops is
\begin{equation}
\begin{split}
& \frac{d \xp}{dt}  =   \vp,   \qquad 
\frac{d \vp }{dt }  = \frac{1}{m_p}\boldsymbol {\mathcal{F}},  \\
& \frac{d r_p}{dt} =  \mathcal{R}, \qquad
\frac{d m_p}{dt}  =  \mathcal{M}, \qquad
 \frac{d T_p}{dt}  =  \mathcal{T}.
\end{split}
\end{equation}

\begin{figure}
  \centering\includegraphics[width=0.8\textwidth]{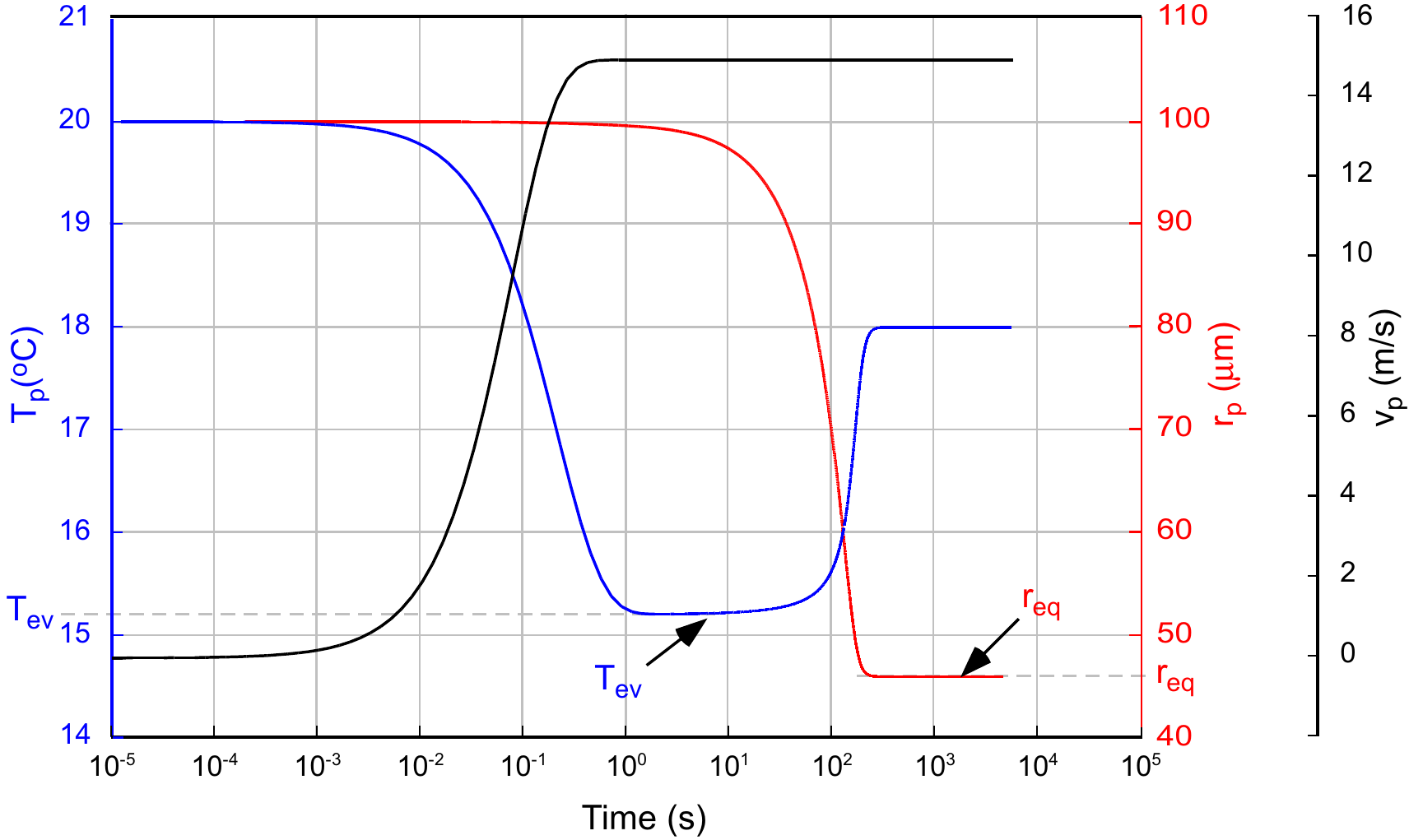}    
	\caption{Time evolution of a water drop’s temperature (blue), radius (red), and velocity (black). The drop is placed at time $t=0$ in air with a 
wind speed of $U_{10}=15$ m s$^{-1}$, a temperature of $T_a=18^oC$ and a relative humidity of $75\%$.The drop has an initial radius of $r_p=100 \mu m$ and initial temperature of $T_p=20^oC$.}
  \label{fig:microphys}
\end{figure}

Figure \ref{fig:microphys} shows the time evolution of a water drop’s temperature (blue), radius (red), and velocity (black) when placed at time $t=0$ in air with a wind speed of $U_{10}=15$ m s$^{-1}$, a temperature of $T_a=18^oC$ and a relative humidity of $75\%$.  The drop is initially at rest and with an initial radius of $r_p=100 \mu m$ and initial temperature of $T_p=20^oC$. From figure   \ref{fig:microphys}, it is apparent that the drop responds quickly to the drag forces imposed on it. It also transfers  sensible heat to the air and rapidly cools down. The latent heat transfer between the air and the drop is slower than both momentum and sensible heat transfers. Indeed, the radius of the drop starts to diminish because of evaporation only after having been exposed to the ambient atmospheric conditions for several seconds. The separation between latent and sensible heat transfers from a spray drop were first noted by \citet{Andreas:1990} who subsequently explored the implication of this phenomena for air-sea spray-mediated heat fluxes \citep[e.g.][]{Andreas:1990, Andreas:1992, Andreas:1995, Andreas:1999, Andreas:2001}. As such, the results of figure \ref{fig:microphys}  are not  new but they will inform results in this paper.

\subsection{Drop conservation and kinetic equation}
Let us now consider a set of drops described by the distribution function
$f(t,\x,\vv,r,m,T)$, which is the number density of drops at time $t$ in the 
phase space $\R^3_{\x} \times \R^3_{\vv} \times [0, +\infty[_r \times [0, +\infty[_m \times [0, +\infty[_T$. It is the number of drops  per $d\x$, per $d\vv$, per $dr$, per $dm$, per $dT$ and has units of $[\textit{(Number of drops)}\  (m)^{-3}\ (m s^{-1})^{-3}\ (m^{-1})\ (kg^{-1})\  (K^{-1})]$.
In other words,  $f(t,\x,\vv,r,m,T)d\x  {d\vv} dr dm dT$ is the number
of drops that at time $t$ are at position $\x\pm d\x$  {(where $d\x=dxdydz$)} with velocity $\vv\pm  {d\vv}$, with a radius $r\pm dr$, with mass $m\pm dm$, and at temperature $T\pm dT$. 

The governing (conservation) equation for this distribution function is
\begin{equation} \label{eq:WB}
\frac{\partial}{\partial t}f
+ \nabv _{\x} \dotv \left( \vv f  \right)
+\nabv _{\vv} \dotv \left(   \frac{\boldsymbol{\mathcal{F}}}{m} f \right) 
+ \frac{\partial}{\partial r} \left( \mathcal{R} f\right) 
+\frac{\partial}{\partial m} \left( \mathcal{M} f \right) 
+\frac{\partial}{\partial T} \left( \mathcal{T} f\right) =0
\end{equation}
This is known as the Williams-Boltzmann equation
\citep{livre-Cox} and can be understood as a conservation law for the number
of drops in the multi-dimensional phase space that represents physical
space $\x$, but also velocity space $\vv$, radius $r$, mass $m$, and
temperature $T$ domains. Indeed, the first two terms of this equation are recognized as the classical conservation law involving the flux divergence. Also, $\nabv_{\vv} \dotv$ is the divergence with respect to the velocity vector. Note that the Eulerian rates of variation of velocity, radius, mass, and
temperature used in~(\ref{eq:WB}) are defined through their Lagrangian
versions~(\ref{eq:Stokes_flow2}),~(\ref{eq:DropRadius}),~(\ref{eq:DropTemp}),~(\ref{eq-M}),
in which the Lagrangian variables $\xp$, $\vp$, $r_p$, $m_p$, $T_p$
are replaced by the Eulerian kinetic variables $\x$, $\vv$, $r$, $m$, and $T$. If the drops do not interact, this change of variable is straightforward. In the case of spray drops, there is a weak interaction because the drops modify the airflow, which in turn can influence the drop transport (see section~\ref{sec:feedback}). In that case, equation \ref{eq:WB} is nonetheless recovered using so called mean-field techniques (see ~\cite{DGR_2008} for a rigorous derivation of a simple example). For particles with strong interactions, such as charged particles submitted to a long range interaction force induced by all the other particles, the derivation of \ref{eq:WB} can be more substantially more complex and the  Eulerian rates of variation require some averaging process. This is not necessary here in the case of sea spray.

Note also that collisions, breakup, coalescence, and nucleation of
drops are normally taken into account by source (and sink) terms in the right-hand side
of~(\ref{eq:WB}). However, since the volume fraction of sea spray is expected to be very small (see the introduction), collision and coalescence are expected to be negligible. Nucleation processes are important in cloud formation but are neglected for sea spray applications. Finally, breakup is expected to be a substantial mechanism of spray formation per se, particularly for large spume drops \citep[e.g][]{Mueller:2009c}, but presumably only for short times after the initial ejection of the spray.
Furthermore, the details of generation of large spume drops at the ocean surface remains poorly resolved. Thus, we consider the spray distribution after the spray is formed and once the drops are no longer fragmenting in the airflow after their ejection from the interface. 
 Overall, source and sink effects are expected to provide minimal contributions to the overall flux balances and can be neglected.

The source of drops can therefore be naturally taken into
  account by boundary conditions. Indeed, equation~(\ref{eq:WB}) is
  set in a moving domain where one boundary is the free sea
  surface. At this boundary, where drops are injected into the airflow, the
  distribution function of incoming drops has to be defined. To that end, let $\x$ be a point on this free surface, $\uu_s(t,\x)$ be
  the velocity of the free surface at this point, and
  $\boldsymbol{n}(t,\x)$ the unit normal vector to this free surface,
  directed toward the air flow. Then, the source of drops coming
  from $\x$ is simply $f(t,\x,\vv,r,m,T)$ for every $\vv$ such that
  $(\vv-\uu_s(t,\x))\dotv \boldsymbol{n}(t,\x)>0$ (i.e. drops whose velocities are directed toward the air flow). This value,
  denoted by $f_{gen}$ thus needs to be known.
  
  Fortunately, because of the interests in assessing spray impacts on Earth’s climate and weather, some efforts have been spent in attempting to evaluate the air-sea flux of spray. In practice, so-called sea spray generation functions, $S(t,\x,r)$, are commonly utilized. They give the number of drops generated at the
  surface per unit surface area, per unit time, and per radius increment. We note however that $S(t,\x,r)$ is not trivial to obtain and nearly all spray generation estimates derive from measurements of drop concentration, $N(t,\x,r)$, at some elevation above the surface. $N(t,\x,r)$ is then the number of drop per unit volume of air, per radius increment, and is a function of space, time, and drop radius. The difference between the flux and the concentration is typically estimated assuming steady state resulting from a balance between gravitational deposition and turbulent suspension. Under these assumptions, the source flux $S(\x,r)$ can then  be inferred from concentration measurements \citep[e.g.][]{Andreas:2010, Veron:2015} and $S(\x,r)= v_{o}(r) N(\x,r)$ where $v_{o}(r)$ is a vertical suspension (or deposition) velocity \citep{Slinn:1980, Andreas:2010}. In that sense, at the interface, $v_{o}=(\vv-\uu_s(t,\x))\dotv \boldsymbol{n}(t,\x)$. However, suspension velocities are likely to be a function of near-surface turbulence and the details of the drop formation process \citep{Mueller:2009c, Veron:2012, Troitskaya:2018a}.

  For the purpose of the model, we assume that the (known) radius
  dependent  ejection velocity is  $\vv_{gen}(r)$\footnote{Note that the ejection velocity can take a more general form in order to include, for example, distributions of ejection angles.}. We also  note that the drops
  generated at the surface have the same density as sea water, and
  therefore their mass $m_{gen}$ is given by $m_{gen}(r) = \frac{4}{3}\pi
  r^3\rho_{sw}$. Likewise, we can also assume that all the drops are ejected with the same sea temperature   $T_{sw}(t,\x)$. Then, dimensional arguments show that the number of drops generated  at  the  surface  per  unit  surface  area,  per  unit  time,  per  velocity,  radius,  mass,  and  temperature increment is
  \begin{equation}  \label{eq-SSGF}
(\vv-\uu_s(t,\x))\dotv \boldsymbol{n}(t,\x) f_{gen}(t,\x,\vv,r,m,T)  =
S \delta_{\vv-\vv_{gen}(r)}\delta_{m-m_{gen}(r)}\delta_{T-T_{sw}(t,\x)},
\end{equation}
where $(\vv-\uu_s(t,\x))\dotv \boldsymbol{n}(t,\x)>0$, and where the ``$\delta$'' symbols denote Dirac delta functions 
  centered in $\vv_{gen}(r)$, $m_{gen}(r)$ and $T_{sw}(t,\x)$, for the velocity, mass, and temperature,  respectively.
Once the $S$ is known, this relation fully
defines the boundary value $f_{gen}$ for equation~(\ref{eq:WB}). 

Finally, note that the Williams-Boltzmann equation~(\ref{eq:WB}) needs to be solved concurrently with the equations for the atmospheric airflow, i.e. the (\textit{Eulerian}) conservation of mass, momentum, and energy respectively:
\begin{align}
\label{eq:Atmo_mass_nofeedback}
\frac{\partial}{\partial t} \rho +      \nabv _{\x} \dotv (\rho \uu)  & =   0, \\
\label{eq:Atmo_momentum_nofeedback}
\frac{\partial}{\partial t} (\rho \uu) + \nabv _{\x} \dotv ( \rho \uu \otimes \uu ) & = \nabv _{\x} \dotv \sigt +  \rho \g, \\
\label{eq:Atmo_energy_nofeedback}
\frac{\partial}{\partial t}  \mathsf{E}  + \nabv _{\x} \dotv  ( \mathsf{E} \uu) &= \nabv _{\x} \dotv {\boldsymbol Q}+ \nabv _{\x} \dotv (\sigt \uu) +\rho \g \dotv \uu,
\end{align}
where $\rho$ is the air density, $\uu$ the air velocity, $\sigt$ atmospheric stress including pressure, $\mathsf{E}$ the total energy density of the air and ${\boldsymbol Q}$ the heat flux.

\section{Airflow including spray effects}
\label{sec:feedback}

The equations \eqref{eq:Atmo_mass_nofeedback}, \eqref{eq:Atmo_momentum_nofeedback}, and \eqref{eq:Atmo_energy_nofeedback} reflect conservation of mass, momentum, and energy for the airflow, but do not include the effects of the spray drops on the atmosphere. Using the kinetic description outlined above, feedback from the drops on
the atmosphere is relatively straightforward to account for. 

\subsection{Conservation of mass}
For example, since $\mathcal{M}dt$ is the mass lost (gained) from a drop during
a small time interval $dt$, the mass  gained (lost) by the fluid during
$dt$, from all the drops at position $\x\pm d\x$ is 
\begin{equation}
-d\x dt\int_{0}^\infty  \int_{0}^\infty  \int_{0}^\infty
\int_{\R^3_{\vv}}  \mathcal{M} f(t,\x,\vv,r,m,T)  {d\vv} dr dm dT.
\end{equation}
Note that the domain considered here now comprises all possible values in the entire phase space $ \R^3_{\vv} \times [0, \infty]_r \times [0, \infty]_m \times [0, \infty]_T$. To lighten the notation, we denote 
\begin{equation}
\int_{0}^\infty  \int_{0}^\infty  \int_{0}^\infty \int_{\R^3_{\vv}} \mathcal{A}(t,\x,\vv,r,m,T)  {d\vv} dr dm dT= \int_{\boldsymbol{ \Omega}_{\mathfrak{X}}} \mathcal{A} d\boldsymbol {\mathfrak{X}}
\end{equation}
where  $\boldsymbol {\mathfrak{X}}=\left(\vv, r, m, T \right)$, and, $ \boldsymbol{ \Omega}_{\mathfrak{X}} = \R^3_{\vv} \times [0, +\infty[_r \times [0, +\infty[_m \times [0, +\infty[_T$.

So, conservation of mass in the air, accounting for the mass exchange between the air and the drops, reads:
\begin{equation} \label{eq-consmass}
\frac{\partial}{\partial t} \rho +      \nabv _{\x} \dotv (\rho \uu)  =    -\prodm
\end{equation}
with \begin{equation}\label{eq-prodm} 
\prodm =   \int_{\boldsymbol{ \Omega}_{\mathfrak{X}}}  \mathcal{M} f d{\boldsymbol{ \mathfrak{X}}},
\end{equation}
the mass production term due to spray drops. In other words, the water vapour transferred from (evaporation) or to (condensation) the spray drops will change the density of the air. In fact, while equations \eqref{eq:Atmo_mass_nofeedback}, \eqref{eq:Atmo_momentum_nofeedback}, and \eqref{eq:Atmo_energy_nofeedback} were pertinent for dry air, it is now necessary to consider moist air, i.e. the (binary) mixture of dry air and water vapour. As such, $\rho$  in equation \eqref{eq-consmass} is the density of moist air and $\rho=\rho_v+\rho_a$, the sum of the water vapour density and the density of dry air. In the remainder of the paper, unless explicitly noted as ``dry'', the air is considered to be moist air.

 
Furthermore, we need an equation for the water vapour density which is used to estimate the rate of variation of both the mass and the temperature of
the drops (see equations~\eqref{eq:DropRadius} and \eqref{eq:DropTemp}). Water vapour advects with the
same velocity as the moist air, plus a small molecular motion $\Uv$ induced by the gradients in water vapour. The equation for the water
vapour density  is then
\begin{equation} \label{equ_water_vapour_with_diffusion1}
\frac{\partial}{\partial t} \rho_v + \nabv _{\x} \dotv (\rho_v \left( \uu+  \Uv \right) )  =   -\prodm.
\end{equation}
The diffusion from molecular motion can further be modeled with a Fick's diffusion law $\Uv=\frac{-1}{q}D_v \nabv _{\x}q$  with diffusivity $D_v$, and where  $q=\rho_v/\rho$ is the specific humidity of air. Thus, the conservation of water vapour reads:
\begin{equation} \label{equ_water_vapour_with_diffusion2}
\frac{\partial}{\partial t} \rho_v +      \nabv _{\x} \dotv (\rho_v \uu ) = \nabv _{\x} \dotv\left( \rho D_v \nabv _{\x}q \right) -\prodm.
\end{equation}
For completeness, equations~\eqref{equ_water_vapour_with_diffusion2} and~\eqref{eq-consmass} yield the mass conservation equation for dry air:
\begin{equation} \label{equ_dry_air_with_diffusion2}
\frac{\partial}{\partial t} \rho_a +      \nabv _{\x} \dotv (\rho_a \uu ) = \nabv _{\x} \dotv\left( \rho D_v \nabv _{\x}(1-q) \right).
\end{equation}
where  $\Ua=\frac{-1}{(1-q)}D_v \nabv _{\x}(1-q)$. Incidentally,  $\rho_a \Ua+\rho_v \Uv=0$ since we consider here a binary mixture with $\rho=\rho_v+\rho_a$.

\subsection{Conservation of momentum}

Equivalently, if ${\boldsymbol F}$ is the force from the fluid on the
drop, the force on the fluid from all drops at position $\x\pm
d\x$ is  $-d\x \int_{\boldsymbol{ \Omega}_{\mathfrak{X}}}  {\boldsymbol F} f  d{\boldsymbol{ \mathfrak{X}}}$. The change of momentum of the air is also due to an increase (decrease) in mass of water vapour coming from evaporating (condensing) drops. The
corresponding rate of variation of momentum is $-d\x \int_{\boldsymbol{ \Omega}_{\mathfrak{X}}}  \vv {\cal M} f  d{\boldsymbol{ \mathfrak{X}}}$.
Then conservation of momentum in the air, accounting for the momentum exchange between the air and the drops, reads:
\begin{equation} \label{eq-consmomentum}
\frac{\partial}{\partial t} (\rho \uu) + \nabv _{\x} \dotv ( \rho \uu \otimes \uu )  = \nabv _{\x} \dotv \sigt +  \rho \g -\produ 
\end{equation}
where 
\begin{equation}
\produ =   \int_{\boldsymbol{ \Omega}_{\mathfrak{X}}}  {\boldsymbol F} f d{\boldsymbol{ \mathfrak{X}}}
+  \int_{\boldsymbol{ \Omega}_{\mathfrak{X}}}  \vv {\cal M} f d{\boldsymbol{ \mathfrak{X}}}
\end{equation}
is the  momentum production term due to the drops. 
The corresponding equation for the air velocity is
\begin{equation}\label{eq-velocity2} 
\begin{split}
  \rho \left( \frac{\partial \uu}{\partial t}  +    \left(\uu \dotv  \nabv _{\x}\right)\uu \right)  +\nabv _{\x} p &= \nabv _{\x} \dotv \taut + \rho \g -\produ +\prodm \uu, \\
  & = \nabv_{\x}\dotv  \taut +  \rho \g 
- \int_{\boldsymbol{ \Omega}_{\mathfrak{X}}}  {\boldsymbol F} f d{\boldsymbol{ \mathfrak{X}}}
-  \int_{\boldsymbol{ \Omega}_{\mathfrak{X}}}  (\vv-\uu)\mathcal{M} f d{\boldsymbol{ \mathfrak{X}}}
\end{split}
\end{equation}
where the stress tensor has been decomposed into its atmospheric pressure and viscous components as $\sigt=-p {\boldsymbol I} + \taut$. Also, the viscous stress tensor is easily obtained from the gradient velocity tensor and its transpose using $\taut =\mu \left( \nabv_{\x} \uu + (\nabv_{\x} \uu)^T -\frac{2}{3} \nabv_{\x} \cdot \uu {\boldsymbol I} \right) $ In developing equation~\eqref{eq-velocity2}, the terms involving molecular diffusive  motions for dry air and water vapour (see equations~\eqref{equ_water_vapour_with_diffusion1} and~\eqref{equ_water_vapour_with_diffusion2}) of order $O(\Ua^2)$ or $O(\Uv^2)$ were neglected.

\subsection{Conservation of energy}

Finally, we seek to estimate the energy exchanged between the drops and the air, noted $\prode$. The energy of a single drop is the sum of its internal and kinetic energy:
\begin{equation}
E_p(t)=m_p(t)e_p(t) + \frac{1}{2}m_p(t)\vp ^2 (t)
\label{equ:drop_E}
\end{equation}
where $e_p$ denotes the drop specific internal energy which is a function of $T_p(t)$. 
So, the rate of change of a spray drop energy is
\begin{align}
\frac{d}{dt}E_p(t)&=\left( e_p(t) +\frac{1}{2}\vp ^2 (t) \right) \frac{d}{dt}m_p(t)+ m_p(t) \left(\frac{d}{dt} e_p(t) + \vp(t)  \dotv \frac{d \vp(t)}{dt} \right) \nonumber \\
&=\left( e_p(t) + \frac{1}{2}\vp ^2 \right) \mathcal{M} + m_p  \frac{d}{dt} e_p(t) +\vp  \dotv \boldsymbol {\mathcal{F}}  \nonumber \\
&=\mathcal{E}(t).
\end{align}
However, the gravity is a force which is external to the closed air/drops system, and its work should not be accounted for in the energy
exchange between the drops and the air. Consequently, to define the rate of variation of energy of the
drop, the work of
$\boldsymbol {\mathcal{F}}$ is replaced by the work of the drag force $\boldsymbol F$ in the
previous relation. Furthermore, the specific internal energy of the drop is simply the specific internal energy of sea water at the temperature of the drop, so $e_p(t)=e_{sw}(T_p(t))$. And since by definition $\frac{d}{dT} e_{sw} = c_{ps}$, then $\frac{d}{dt} e_{sw}(T_p(t))=c_{ps} \frac{dT_p(t)}{dt}=c_{ps} \mathcal{T}$ in which $c_{ps}$ is considered constant.  In fact, equation~\eqref{eq:DropTempPruppacher} can now be understood as the rate of change of a drop's internal energy. Finally,
\begin{equation}  \label{eq-defprode}
\mathcal{E}(t) =  e_{sw}(T_p)  \mathcal{M} + m_p  c_{ps}\mathcal{T} +\vp  \dotv \boldsymbol F + \frac{1}{2}\mathcal{M} \vp ^2 .
\end{equation}
In the equation above, one can see that changes in internal energy of a drop may arise from both a change of mass and a change of temperature, while the change of kinetic energy results from the work of friction forces and a change in drop mass. As for the mass and momentum conservation equations above, the energy production term from all the drops is
\begin{equation}  \label{eq-prode}
  \prode =  \int_{\boldsymbol{ \Omega}_{\mathfrak{X}}}  \mathcal{E} f d{\boldsymbol{ \mathfrak{X}}}
\end{equation}
in which $\mathcal{E}$
  has to be understood as an Eulerian quantity derived
  from its Lagrangian definition~(\ref{eq-defprode}). In other words, equation~\eqref{eq-prode} reads:
\begin{equation} \label{eq-prode2}
\prode  
= \int_{\boldsymbol{ \Omega}_{\mathfrak{X}}}  e_{sw}(T)  \mathcal{M}
  f  d\boldsymbol {\mathfrak{X}} 
 + \int_{\boldsymbol{ \Omega}_{\mathfrak{X}}} m c_{ps}   \mathcal{T}  f  d\boldsymbol {\mathfrak{X}} 
 +\int_{\boldsymbol{ \Omega}_{\mathfrak{X}}} \vv  \dotv \boldsymbol {F} f  d\boldsymbol
{\mathfrak{X}}  +\int_{\boldsymbol{ \Omega}_{\mathfrak{X}}} \frac{1}{2} \vv ^2\mathcal{M}f  d\boldsymbol
{\mathfrak{X}}.
\end{equation}

We can now develop the energy conservation for the air, accounting for the input from the spray. However, at this stage is is necessary to separate the moist air energy density $\mathsf{E}$ into energy density for dry air $\mathsf{E}_a$, and that of the water vapour, $\mathsf{E}_v$. With spray effects included, equation~\eqref{eq:Atmo_energy_nofeedback} reads:
\begin{align} \label{eq-consenergy_moist}
\frac{\partial}{\partial t}  \left(\mathsf{E}_a +\mathsf{E}_v \right)  +& \nabv _{\x} \dotv
(\left(\mathsf{E_a} +\mathsf{E_v} \right) \uu)  + \nabv _{\x} \dotv
\left(\mathsf{E}_a \Ua +\mathsf{E}_v\Uv \right) \nonumber \\
&= \nabv _{\x} \dotv {\boldsymbol Q} + \nabv _{\x} \dotv (\taut \uu) - \nabv _{\x} \dotv \left( (\uu+\Ua)p_a+(\uu+\Uv)p_v \right) \nonumber \\ & \quad +\rho \g \dotv \uu -\prode
\end{align}
 where $p_a$ is the partial pressure for dry air, and $p_v$ the water vapour partial pressure. The separation between water vapour and dry air is useful here because we wish to keep the rate of work done by the partial pressures. We note here that neglecting $\Uv p_v$ and $\Ua p_a$  would allow us to simply use $p=p_v+p_a$ and incorporate the pressure in the stress tensor of the (moist) air as in equation~\eqref{eq-velocity2}. However, it is convenient to keep the partial pressures explicitly out of the stress tensor so that expressions for the enthalpy of the dry air and water vapour can be uncovered. For example, as for the drop, the energy density of the water vapour can be decomposed into internal and kinetic energy:
 \begin{align} \label{eq-energy_water_vapour}
\mathsf{E}_v=\rho_v e_v+ \frac{1}{2}\rho_v |\uu|^2
\end{align}
where $e_v$ is the specific internal energy for water vapour, and where the contribution of the diffusion velocity to the kinetic energy can be neglected. By definition, the specific enthalpy for the water vapour is $h_v=e_v+p_v/\rho_v$, and thus 
 \begin{align} \label{eq-enthalpy_water_vapour}
\mathsf{E}_v+p_v=\rho_v h_v+ \frac{1}{2}\rho_v |\uu|^2.
\end{align}
Similarly,
 \begin{align} \label{eq-enthalpy_dry_air}
\mathsf{E}_a+p_a=\rho_a h_a+ \frac{1}{2}\rho_a |\uu|^2.
\end{align}
with the specific enthalpy of dry air $h_a=e_a+p_a/\rho_a$, and $e_a$ the specific internal energy for dry air.
Therefore, equation~\eqref{eq-consenergy_moist} can be written as:
\begin{align} \label{eq-consenergy_moist2}
\frac{\partial}{\partial t}  \left(\rho_a e_a+\rho_v e_v \right)  &+ \nabv _{\x} \dotv
(\left(\rho_a h_a+\rho_v h_v \right) \uu)  \nonumber \\
&+ \nabv _{\x} \dotv \left(\rho_a h_a \Ua  +\rho_v h_v \Uv  \right)\nonumber \\
&+\frac{\partial}{\partial t}  \left(\frac{1}{2}\rho  |\uu|^2 \right)
+ \nabv _{\x} \dotv \left(\frac{1}{2}\rho |\uu|^2 \uu  \right)\nonumber \\
&= \nabv _{\x} \dotv {\boldsymbol Q} + \nabv _{\x} \dotv (\taut \uu) +\rho \g \dotv \uu -\prode
\end{align}

Using  equations~\eqref{equ_water_vapour_with_diffusion2}, ~\eqref{eq-consmomentum}, and~\eqref{eq-consenergy_moist2}, we find the equation governing $e$, the specific internal energy of moist air:
\begin{align} \label{eq-energy_intern_moist_air}
\frac{\partial}{\partial t}  \left(\rho e \right)  &+ \nabv _{\x} \dotv
\left( \rho e \uu \right)  \nonumber \\
&= \nabv _{\x} \dotv {\boldsymbol Q} + \taut:\nabv_{\x} \uu  -p \nabv_{\x} \dotv \uu \nonumber \\
&- \nabv _{\x} \dotv {\boldsymbol J} \nonumber \\
&+ \produ \dotv   \uu  - \frac{1}{2} |\uu|^2 \prodm -\prode
\end{align}
where $\rho e=\rho_a e_a+\rho_v e_v$, and 
\begin{align}\label{eq-Kiva}
{\boldsymbol J} &=\rho_a h_a \Ua +\rho_v h_v \Uv \nonumber \\
&=-\rho D_v\left( h_a\nabv _{\x}(1-q) + h_v\nabv _{\x}q\right)
\end{align}
is a flux of enthalpy  due to small molecular motions.

For modeling purposes, it is generally practical to look at the
temperature  of the carrier flow $T_a$. 
With tedious but simple algebra, the energy conservation equation~\eqref{eq-energy_intern_moist_air} can be
written as
\begin{align}\label{eq-Tair} 
\rho c_\textrm{v}  \left( \frac{\partial T_a}{\partial t}  +   \uu \dotv \nabv_{\x} T_a \right)  
 = &  \nabv_{\x}\dotv (\kappa \nabv_{\x} T_a) +  \taut:\nabv_{\x} \uu  -p \nabv_{\x} \dotv \uu \nonumber  \nonumber \\
 & - \nabv _{\x} \dotv {\boldsymbol J} + e_a \nabv _{\x} \dotv (\rho_a \Ua) + e_v \nabv _{\x} \dotv (\rho_v \Uv) \nonumber \\
& + \produ \dotv   \uu  - \frac{1}{2} |\uu|^2  \prodm -\prode +e_v \prodm 
\end{align}
where  we have taken the heat
flux as given by the Fourier law ${\boldsymbol Q}  = \kappa \nabv_{\x} T_a$ and where $c_\textrm{v}$ is the specific heat of air at constant volume, derived from the corresponding specific heat of water vapour and dry air using:
\begin{equation} \label{eq-Tair2} 
c_\textrm{v}=q c_{\textrm{v},v}+(1-q)c_{\textrm{v},a}
\end{equation}
Also, in equation~\eqref{eq-Tair}, $e_v$ and $e_a$ are taken at the air temperature $T_a$.

Expanding the spray production terms leads to:
\begin{equation}\label{eq-Tair3} 
\begin{split}
\rho c_\textrm{v}  \left( \frac{\partial T_a}{\partial t}  +   \uu \dotv \nabv_{\x} T_a \right) 
    = &   \nabv_{\x}\dotv (\kappa \nabv_{\x} T_a) +  \taut:\nabv_{\x} \uu  -p \nabv_{\x} \dotv \uu  \\
& \quad - \nabv _{\x} \dotv {\boldsymbol J} + e_a \nabv _{\x} \dotv (\rho_a \Ua) + e_v \nabv _{\x} \dotv (\rho_v \Uv) \\
&\quad  -\int_{\boldsymbol{ \Omega}_{\mathfrak{X}}} (\vv-\uu)  \dotv \boldsymbol {{F}} f  d\boldsymbol
{\mathfrak{X}} \\
&\quad - \int_{\boldsymbol{ \Omega}_{\mathfrak{X}}}    \frac{1}{2} |\vv-\uu|^2  {\cal M}
f d{\boldsymbol{ \mathfrak{X}}}\\
& \quad  - \int_{\boldsymbol{ \Omega}_{\mathfrak{X}}} m c_{ps}   \mathcal{T}  f  d\boldsymbol {\mathfrak{X}} \\
& \quad  + \int_{\boldsymbol{ \Omega}_{\mathfrak{X}}} \left( e_v(T_a)-e_{sw}(T)\right)   \mathcal{M}  f  d\boldsymbol {\mathfrak{X}}
\end{split}
\end{equation}


In the equation above, the spray feedback terms are on the right hand side and evidently contain $f$ and its integrals. The first
 three feedback term represents the heat gained by the air due to the work of the frictional drag force exerted by the drops on the air, a loss of  kinetic energy of the drops, and the exchange of sensible heat between the air and the drops due to the variation of the temperature of the drops. The last feedback term represents a transfer in internal energy from the drops to water vapour as drops evaporate.

Also, we develop the term resulting from molecular diffusion,  
\begin{align}
\mathcal{D} &= - \nabv _{\x} \dotv {\boldsymbol J} + e_a \nabv _{\x} \dotv (\rho_a \Ua) + e_v \nabv _{\x} \dotv (\rho_v \Uv) \nonumber \\
&= R T_a \left( \frac{1}{M_w} - \frac{1}{M_a}\right)\nabv _{\x} \dotv \left( \rho  D_v \nabv _{\x} q\right)\nonumber \\
&\qquad+ \left( c_{\textrm{p},v} -c_{\textrm{p},a} \right) \rho D_v \nabv _{\x}q \dotv  \nabv _{\x}T_a  
\end{align}
where $c_{\textrm{p},v}$ and $c_{\textrm{p},a}$ are respectively the specific heat at constant pressure for water vapour and dry air. Finally,
\begin{align}
 e_{sw}(T)&= c_{ps}(T-T_{ref})+e_{sw}(T_{ref}) \quad \text{and,}\nonumber \\
 e_v(T_a)&= c_{\textrm{v},v} (T_a-T_{ref})-\frac{R}{M_w}T_{ref}+L_v
\end{align}
where $T_{ref}$ is a reference temperature, generally taken as $T_{ref}=0^oC$, such that the latent heat of vaporization, $L_v$, is the enthalpy difference between water vapour and liquid water at $T_{ref}$. To simplify the notation, we write the last feedback term in~\eqref{eq-Tair3} as:
\begin{equation}
\Upsilon=\int_{\boldsymbol{ \Omega}_{\mathfrak{X}}} \left( e_v(T_a)-e_{sw}(T)\right)   \mathcal{M}  f  d\boldsymbol {\mathfrak{X}}
\end{equation}

\section{Dimensional analysis}
In this section, we attempt to simplify the system given by William-Boltzman equation (equation~\eqref{eq:WB}),  with the atmospheric flow conservation equations which all include the contributions from the spray (equations  \eqref{eq-consmass}, \eqref{eq-velocity2}, and  \eqref{eq-Tair3}).

\subsection{Characteristic quantities and dimensionless variables}
\label{subsubsec:char}

Here, we take $\rchar$ as a characteristic radius of a population of drops. Their corresponding vertical settling velocity is denoted by $\vchars$.  
The space scale $\xchar$ is a typical transport scale for these drops which we choose as the significant wave height \citep{Andreas:2004}. Thus, the time scale $\tchar = \xchar/\vchars$  roughly corresponds to a typical average time for a drop to fall from the significant wave height to the mean water level. \citet{Andreas:2004} considers $\tchar$ as a viable proxy for the drop time of flight, but we emphasize here that the details of the ejection mechanism such as the initial velocity and angle of ejection of the drops are not accounted for here. There is some evidence that large drops ejected near the wave crest may fall back near the crest and not at the mean water level \citep{Mueller:2014a}. And the ejection velocity, particularly for small, bubble-generated drops, can be significant \citep{Blanchard:1989, Spiel:1995, Spiel:1997,Spiel:1998}. Nevertheless, for the purpose of the non-dimensional scaling, $\tchar$ is an adequate estimate for the drop time of flight. The characteristic temperature of the drops is naturally taken as the sea temperature and the characteristic mass drop is therefore $\mchar = \rhosw \frac{4}{3}\pi \rchar ^3$. Finally, the corresponding characteristic values $\mathcal{F}_o$, $\Rchar$,
$\Mchar$, and $\TTchar$ of the the rates of variation $\boldsymbol {\mathcal{F}}$,
$\mathcal{R}$, $\mathcal{M}$, $\mathcal{T}$ of the drop momentum, radius,
mass and temperature, respectively, are obtained by using
formulas~\eqref{eq:drag_law},~\eqref{eq:DropRadius},~\eqref{eq:DropTemp}, and~\eqref{eq-M}.  Because sea spray is made of spray drops that are disperse in size (ranging here from $O(1)\; \mu m$ to $O(1)\; mm$), defining the characteristic value $\fchar$ of the
distribution function $f$ is delicate. However, we shall see later that $\fchar$ appears in all spray feedback terms and thus naturally falls out of a term-by-term dimensional comparison.

For the air flow, we take the same space and time scales. However, since air flow and spray drops  can have very different velocities, especially for large spume drops with substantial mass and inertia, we
use a velocity scale $\vchar$ for the air flow as a typical wind speed, possibly different from $\vchars$. An obvious choice for $\vchar$ is the mean wind speed measured at 10m height above the surface, $U_{10}$. The characteristic
air mass density is $\rhochar$, and its characteristic
temperature is $\Tchar$ which is the same as the characteristic sea temperature. Derived characteristic values of viscosity $\muchar$
and heat transfer $\kappachar$ coefficients can be easily obtained. 

In the following, the corresponding dimensionless quantities will be denoted with the superscript $'$, and defined by $[.]'=[.]/[.]_o$; for instance $\ts = t/\tchar$ for the time variable. 



\subsection{Dimensionless Williams-Boltzmann equation}

The Williams-Boltzmann
equation~\eqref{eq:WB} can be non-dimensionalized to
\begin{equation} \label{eq:WBaf} 
\begin{split}
& \frac{\partial}{\partial \ts} \fs
 +  \nabv _{\xs} \dotv \left(\fs \vs  \right)  
 + \frac{\tchar}{\tau_v} \nabv _{\vs} \dotv 
\left(  \frac{\fs}{\ms} \mathcal{F}' \right) \\
& + \frac{\tchar}{\tauR}\frac{\partial}{\partial \rs} \left( \Rs
  \fs\right) 
 +\frac{\tchar}{\taum}\frac{\partial}{\partial \ms} \left(  \Ms
  \fs \right) 
 +\frac{\tchar}{\tauT}
\frac{\partial}{\partial \Ts} \left(
\TTs \fs\right)
=0,
\end{split}
\end{equation}
where  the relaxation times $\tau_v$, $\tauR$, $\taum$, and $\tauT$ are characteristic
times for the variation of the momentum, radius, mass, and temperature of the drops, defined by
\begin{equation}\label{eq-relaxation_times} 
\tau_v=\frac{\mchar \vchars}{\mathcal{F}_o}, \qquad
  \tauR = \frac{\rchar}{\Rchar}, \qquad 
\taum  = \frac{\mchar}{\Mchar}, \qquad
  \tauT = \frac{\Tchar}{\TTchar}.
\end{equation}

We note that the momentum term can further be decomposed into its gravitational and drag components such that
\begin{equation} 
\frac{\tchar}{\tau_v} \nabv _{\vs} \dotv \left(  \frac{\fs}{\ms} \mathcal{F}' \right)
=\nabv _{\vs} \dotv 
\left(  \frac{\tchar }{\taugrav}   \gs \fs  
        +\frac{\tchar}{\tauIchar}  \frac{1}{\tauIs}\left(\frac{1}{\ratiov}\us -\vs\right)\fs\right)
\end{equation}
where  $\ratiov=\vchars/\vchar$ is the ratio of
the characteristic drop and air flow velocities, and where the dimensionless
gravity vector is $\gs =\g/g = -e_z$, with $g=|\g|$ the gravity
magnitude.
Then, $\taugrav=\vchars/g$ is the characteristic drop time scale due to gravity and $\tauIchar=2r_o^2\rho_{sw}/\left(9\muchar \Xi_o\right)$ is the characteristic drop response time to the drag force, i.e   $\tauI=\tauIchar \tauIs$ with  $\tauIs=r'^2/(\mu' \Xi')$. 

\begin{figure}
  \centering\includegraphics[width=0.7\textwidth]{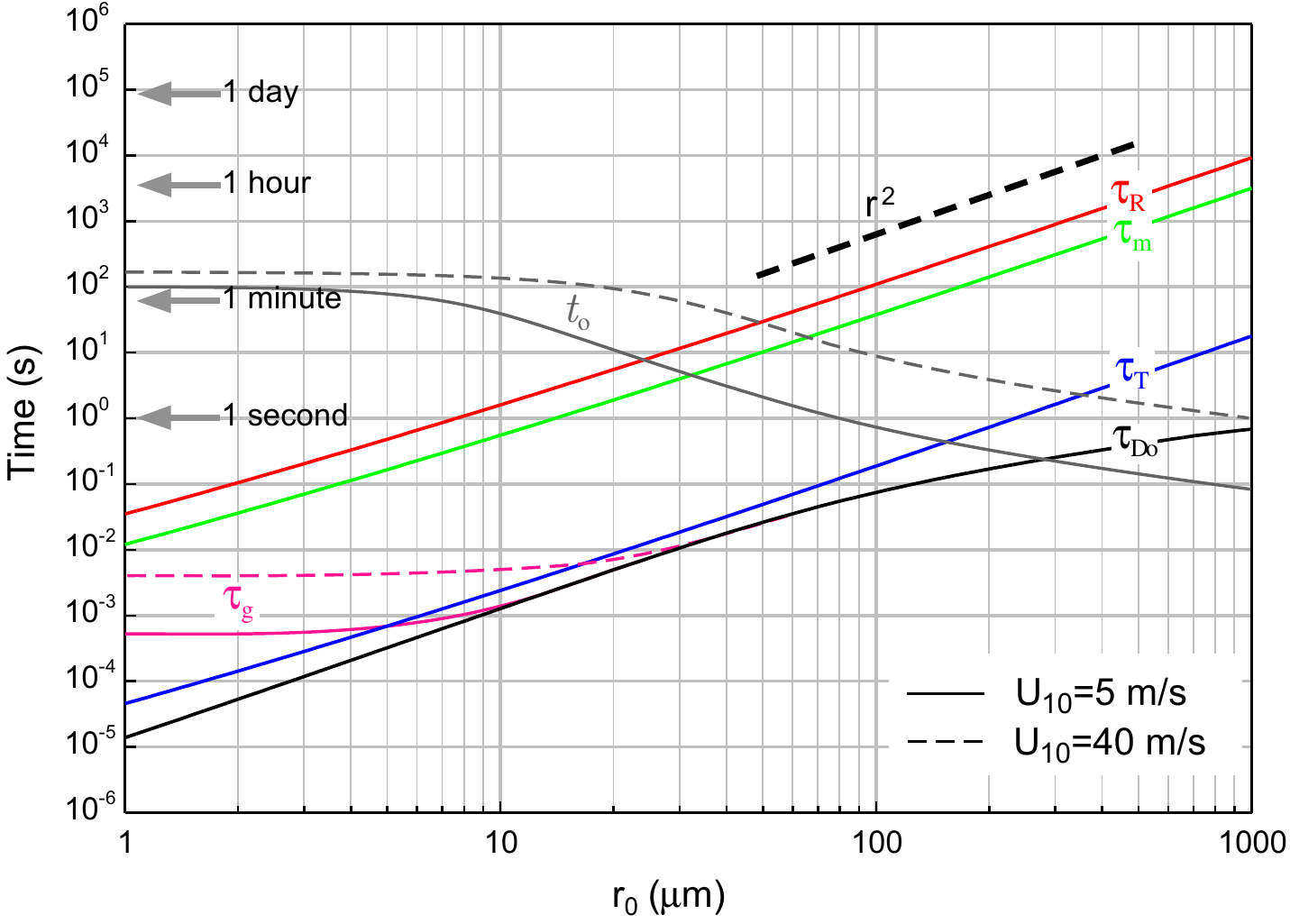}    
	\caption{Characteristic times of variation of the drag force
          $\tau_{Do}$, radius $\tau_R$, mass $\tau_m$, and temperature
          $\tau_T$, of the drop as a function of the characteristic
          radius $r_0$ and for water temperature of $T_a=18^oC$,
          $T_p=20^oC$, and a $75\%$ relative humidity. Also, the
          characteristic times $\taugrav$ and $\tchar$ are shown for
          wind speeds of $U_{10}=5$ m s$^{-1}$ with a significant wave
          height of $\xchar=0.5$ m (solid lines), and $U_{10}=40$ m
          s$^{-1}$ with a significant wave height of $\xchar=6.4$ m
          (dashed lines). }
  \label{fig:Figure1}
\end{figure}

Figure \ref{fig:Figure1} shows the characteristic times of variation of the drag force $\tau_{Do}$, radius $\tau_R$, mass $\tau_m$, and temperature $\tau_T$, of the drop as a function of the characteristic radius $r_0$. These timescales are plotted here for specific ambient conditions, but they do vary (albeit weakly), in particular with relative humidity \citep[see][]{Andreas:1990}. They are, however, strong functions of the drop radius. Indeed, $\tau_{Do}$ is proportional to the drag on the drop, i.e. its cross sectional area, and $\tau_R$, $\tau_m$, and  $\tau_T$ are all controlled by molecular exchanges through the drops surface area. Thus, all these time scales are proportional to $r_o^2$.
From figure \ref{fig:Figure1}, one can easily see that small drops transfer sensible heat faster than larger
ones do. Also, $\tau_T\ll\tau_R$ and the temperature of a drop will change and adapt to ambient atmospheric conditions much faster (approximately 1000 times) than its radius will.

We note here that in the absence of wind or other sources of airflow fluctuations and turbulence, the terminal fall velocity of a drop, $\textrm{v}_t(r_o)$,  is such that  $\textrm{v}_t/g=\tauIchar$, and $\textrm{v}_t(r_o)$ is a function of the drop radius only. However, when the airflow is turbulent, the terminal velocity $\textrm{v}_t$ and the settling (or deposition) velocity  $\vchars$, of small drops in particular, can be substantially different. Indeed, small drops are affected by the turbulence in the airflow such that $\textrm{v}_t$ does not adequately represent the actual velocity at which these drops effectively settle \citep[e.g.][]{Slinn:1980, Andreas:2010}. Here, we adopt a formulation found in \citet{Smith:1993} and originally proposed by \citet{Carruthers:1986}
\begin{equation}
\vchars=\frac{\textrm{v}_t}{1-\textrm{exp} \left(\frac{\textrm{v}_t}{C_D U_{10}}\right)},
\end{equation}
where $C_D\sim O(10^{-3})$ is the air-sea aerodynamic drag coefficient. 
Thus $\taugrav$ is function of both the drop radius and the wind speed. It is plotted on figure \ref{fig:Figure1} for $U_{10}=5$ m s$^{-1}$ with a significant wave height estimated at $\xchar=0.5$ m (solid line), and for $U_{10}=40$ m s$^{-1}$ with a significant wave height estimated at $\xchar=6.4$ m (dashed line). 

It is worthwhile to compare the relaxation times with $\tchar$, the atmospheric residence time of spray drops. For convenience, we have plotted $\tchar$ on figure \ref{fig:Figure1}, and we note that, $\tchar$ also depends on the drop settling velocity and thus on the wind speed. For the remainder of the analysis, it is useful to introduce the following dimensionless numbers
\begin{equation}  \label{eq-nodim}
\agravs = \frac{\tchar}{\taugrav}, \quad
\aI = \frac{\tchar}{\tauIchar}, \quad
\aR = \frac{\tchar}{\tauR}, \quad
\am = \frac{\tchar}{\taum}, \quad
\aT = \frac{\tchar}{\tauT},
\end{equation}

such that the nondimensionalized Williams-Boltzmann equation can be written as:
\begin{equation} \label{eq:WBaf} 
\begin{split}
& \frac{\partial}{\partial \ts} \fs
 +  \nabv _{\xs} \dotv \left(\fs \vs  \right)  \\
& + \nabv _{\vs} \dotv 
\left(  \agravs
  \gs \fs  
        +\aI  \frac{1}{\tauIs}\left(\frac{1}{\ratiov}\us -\vs\right)\fs\right) \\
& + \aR\frac{\partial}{\partial \rs} \left( \Rs
  \fs\right) \\
& +\am\frac{\partial}{\partial \ms} \left(  \Ms
  \fs \right) \\
& +\aT
\frac{\partial}{\partial \Ts} \left(
\TTs \fs\right)
=0.
\end{split}
\end{equation}

\begin{figure}
  \centering\includegraphics[width=0.7\textwidth]{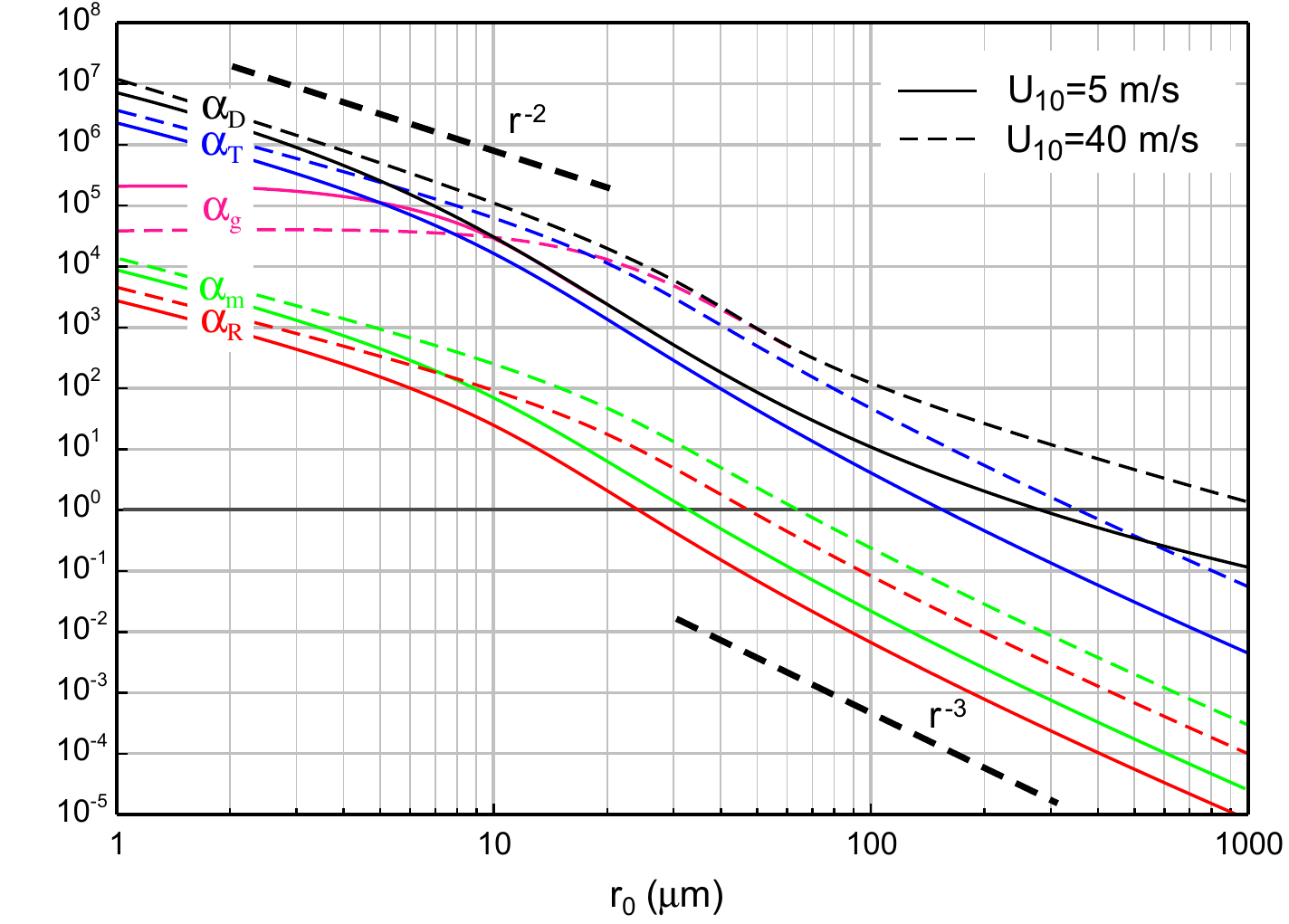}    
	\caption{Non-dimensional relaxation times for the drag force on the drop $\aI$, the drop temperature $\aT$, the drop mass $\am$, and the drop radius $\aR$. Solid lines indicate $U_{10}=5$ m s$^{-1}$ with a significant wave height estimated at $\xchar=0.5$ m, and dashed lines  indicate $U_{10}=40$ m s$^{-1}$ with a significant wave height estimated at $\xchar=6.4$ m.}
  \label{fig:alpha}
\end{figure}

Figure \ref{fig:alpha} shows these non-dimensional \textit{alphas} which are the ratios of the drops residence time $\tchar$, to  the relaxation times $\tau_v$, $\tauR$, $\taum$, and $\tauT$. They are plotted as a function of $r_o$ and for a wind speed of $U_{10}=5$ m s$^{-1}$ (solid lines) and  $U_{10}=40$ m s$^{-1}$ (dashed lines). When these ratios are above 1, this simply indicates that the drops remain suspended in the airflow long enough to exchange their momentum, sensible heat, or water (vapour) mass with the surrounding atmosphere. For example, from figure \ref{fig:alpha}, we deduce that at $U_{10}=5$ m s$^{-1}$, drops with $r_o < 150\; \mu$m will exchange momentum and sensible heat with the atmosphere, while only drops with $r_o < 30\; \mu$m will have the opportunity to exchange latent heat with the air, thereby transferring water vapour to the atmosphere.

\subsection{Dimensionless equations for the airflow including feedbacks}
In this section, we non-dimensionalize the equations for the airflow including spray feedback effects.

\subsubsection{Conservation of mass} 
The dimensionless mass conservation equation for the air is
\begin{equation}\label{eq:Mass_Admin}  
\frac{\partial}{\partial t'} \rho' +      
\frac{1}{\ratiov}\nabv _{\x'} \dotv (\rho '\uu')  
=-\am\chi
\int_{\boldsymbol{ \Omega}_{\mathfrak{X'}}}  \mathcal{M'} f'  d\boldsymbol {\mathfrak{X'}},
\end{equation}
where $\chi = \mchar \fchar \vchars^3\rchar\Tchar\mchar/\rhochar$. 
Physically, $\chi$ is a mass fraction, i.e. the mass of the drops represented by $\fchar$, divided by the mass of a characteristic volume of air.

\subsubsection{Airflow velocity}
In non-dimensional variables, equation~(\ref{eq-velocity2}) reduces to  
\begin{equation}\label{eq:Vel_Admin}
\begin{split}
 \rhos \left(\frac{\partial \us   }{\partial \ts}  
+  \frac{1}{\ratiov}(\us \dotv \nabv_{\xs}) \us \right) 
+  \frac{1}{\ratiov}\frac{1}{\gamma \Ma^2}\nabv_{\xs} \ps   
= &\frac{1}{\ratiov}\frac{1}{\Reyn}\nabv_{\xs}\dotv (\mus (\nabv_{\xs} \us + \nabv_{\xs}\us^T -
\frac{2}{3}\nabv_{\xs}\us {\boldsymbol I})) \\
 & +  \frac{1}{\ratiov}\agrav \rhos \gs \\
&  
-\aI\chi  
\int_{\boldsymbol{ \Omega}_{\mathfrak{X'}}} \frac{\ms}{\tauIs}(\us -\ratiov\vs)
 \fs  d\boldsymbol {\mathfrak{X'}}\\
&
-\am \chi 
\int_{\boldsymbol{ \Omega}_{\mathfrak{X'}}}  (\ratiov\vs-\us)\mathcal{\Ms} \fs d{\boldsymbol{ \mathfrak{X'}}},
\end{split}
\end{equation}
where the nondimensional number in front of the pressure gradient contains the
Mach number of the air flow defined by $\Ma = \vchar/\cchar$ with $\cchar = \sqrt{\gamma \pchar/\rhochar}$ the sound speed in
air, $\gamma = c_\textrm{p}/c_\textrm{v}$ the
ratio of specific heats of air, and $\pchar = \rhochar R \Tchar$  the characteristic value for the
atmospheric pressure. The nondimensional number in front of
the shear stress is the inverse of the Reynolds number of the air flow 
$\Reyn = \rhochar\vchar\xchar/\muchar$, where $\muchar = \mu(\Tchar)$ is a characteristic
value for the viscosity of air. The non-dimensional
number in the gravity term contains the Froude number ${\rm Fr} =
\vchar/\sqrt{\xchar g} $.
The other dimensionless
numbers have been defined in previous sections.

It is possible to substantially simplify the equation above. Indeed, we can see from  figure \ref{fig:alpha} that $\alpha_m\ll\alpha_D$ for all drop radii. Therefore, among the spray feedback terms, only that related to the drag force on the fluid from the drops remains.

\subsubsection{Airflow temperature}

The same procedure leads to the nondimensional form of the temperature equation (equation \eqref{eq-Tair3}):
\begin{equation}\label{eq:Temp_Admin}
\begin{split}
&  %
\rhos \left( \frac{\partial \Tas}{\partial \ts}
+ 
 \frac{\us}{\ratiov}
\dotv \nabv_{\xs} \Tas \right)
\\
&  = 
 \frac{1}{\ratiov}\frac{\gamma}{\Pe} \nabv_{\xs}\dotv (\kappas
\nabv_{\xs} \Tas) \\
&\quad  
+ 
 \frac{1}{\ratiov}\gamma(\gamma
-1)\frac{\Ma^2}{\Reyn}
\mus\left(\nabv_{\xs} \us + \nabv_{\xs} \us^T - \frac{2}{3}\nabv_{\xs}\dotv
\us I\right) : \nabv_{\xs} \us \\
& \quad 
- \frac{1}{\ratiov}(\gamma-1) \ps \nabv_{\xs}\cdot \us \\
& \quad 
+ \mathcal{D}' \\
& \quad + \gamma(\gamma-1)\Ma^2\aI\chi
\int_{\boldsymbol{ \Omega}_{\mathfrak{X'}}}  
\frac{\ms}{\tauIs}|\us-\ratiov \vs|^2 \fs  d\boldsymbol  {\mathfrak{X'}},\\
& \quad - \gamma(\gamma-1)\Ma^2\am \chi
\int_{\boldsymbol{ \Omega}_{\mathfrak{X}}}  \left(
  \frac{1}{2} |\ratiov\vs-\us|^2 \right) \Ms \fs d{\boldsymbol{\mathfrak{X'}}}\\
& \quad - \frac{c_{ps}}{c_\textrm{v}}\aT\chi
\int_{\boldsymbol{ \Omega}_{\mathfrak{X'}}} \ms   \TTs  \fs
d\boldsymbol {\mathfrak{X'}}\\
& \quad + \Upsilon'
\end{split}
\end{equation}
where the dimensionless number in front of
the heat flux contains the Peclet number  for heat $\Pe = c_\textrm{p}\xchar\vchar\rhochar/\kappachar$, with $c_\textrm{p}$ the specific heat of air at constant pressure. 

The non dimensional diffusion term reads:
\begin{align}
\mathcal{D}'&=
\frac{1}{\ratiov} \frac{1}{\Pem} \frac{R}{c_\textrm{v}} q_o \Tas \left( \frac{1}{M_w} - \frac{1}{M_a}\right)  \nabv _{\xs} \dotv \left( \rhos  \nabv _{\xs} \qs \right) \nonumber \\
& \quad 
+ \frac{1}{\ratiov} \frac{1}{\Pem} q_o\frac{c_{\textrm{p},v}-c_{\textrm{p},a}}{c_\textrm{v}} \rhos \nabv _{\xs} \qs \dotv \nabv _{\xs} \Tas
\end{align}
where $\Pem$ is the Peclet number for mass transfer of water vapour $\Pem =\xchar\vchar/D_{v}$ and $q_o$ is a characteristic specific humidity. Finally, the non-dimensional internal energy term:
\begin{align}\label{eq:internal_energy_Admin}
\Upsilon'&=
\frac{c_{\textrm{v},v}}{c_\textrm{v}} \am \chi
\int_{\boldsymbol{ \Omega}_{\mathfrak{X}}}  \left(
  \Tas-\frac{T_{ref}}{\Tchar} \right) \Ms \fs d{\boldsymbol{ \mathfrak{X'}}} \nonumber \\
& \quad 
+ \frac{L_v}{\Tchar c_\textrm{v}} \am \chi
\int_{\boldsymbol{ \Omega}_{\mathfrak{X}}} \Ms \fs d{\boldsymbol{ \mathfrak{X'}}} \nonumber \\
& \quad 
- \frac{R T_{ref}}{M_w \Tchar c_\textrm{v}} \am \chi
\int_{\boldsymbol{ \Omega}_{\mathfrak{X}}} \Ms \fs d{\boldsymbol{ \mathfrak{X'}}}  \nonumber \\
& \quad 
- \frac{c_{ps}}{ c_\textrm{v}} \am \chi
\int_{\boldsymbol{ \Omega}_{\mathfrak{X}}}  \left(
  \Ts-\frac{T_{ref}}{\Tchar} \right)\Ms \fs d{\boldsymbol{ \mathfrak{X'}}} \nonumber \\
  & \quad 
- \frac{e_{sw}(T_{ref})}{ \Tchar} \am \chi
\int_{\boldsymbol{ \Omega}_{\mathfrak{X}}} \Ms \fs d{\boldsymbol{ \mathfrak{X'}}}
\end{align}

Again, using figure~\ref{fig:alpha}, it is possible to see that several feedback terms can be neglected. With the typical values for $c_{ps}$, $c_\textrm{v}$, $c_\textrm{p}$, $c_{\textrm{p},v}$, $c_{\textrm{p},a}$, $\gamma$, and considering a low Mach number (see below) and values of $T$ and $\Tchar$ between 0 and about $40^oC$, only two feedback terms remain. They are the spray feedback due to sensible heat variations from the drop (second to last term in in~\eqref{eq:Temp_Admin}) and the latent heat transfer from the drops to the water vapour (second term on the right-hand side of~\eqref{eq:internal_energy_Admin}).

\section{Discussion}
\label{sec:Discussion}
\subsection{Spray feedback estimates}

Evidently, while individual drops may have sufficient time to transfer a particular property to the atmosphere, the effect on the airflow (feedback), if any, depends on the aggregated drop effects, and thus on the drop size distribution and its integrals. Below, we present an example application of the model in which we examine the spray effects on the airflow. However, a comprehensive study of the spray effects on the atmosphere over a wide range of environmental conditions is beyond the scope of this paper. Here, we simply illustrate the capability of the model with an example which corresponds to realistic marine weather conditions, yet not as extreme as storm events during which sea spray is known to be ubiquitous, and for which the present model is certainly expect to work well.

To that effect, we consider the simplified equations obtained from the dimensional analysis in the section above. In dimensional form, they are:

\begin{equation} \label{eq-consmass_dim}
\frac{\partial}{\partial t} \rho +      \nabv _{\x} \dotv (\rho \uu)  =    -\int_{\boldsymbol{ \Omega}_{\mathfrak{X}}}  \mathcal{M} f d{\boldsymbol{ \mathfrak{X}}},
\end{equation}

\begin{equation}\label{eq-q_dim} 
  \frac{\partial}{\partial t} q +       \uu\dotv \nabv _{\x}  q  =    -\frac{1-q}{\rho}\int_{\boldsymbol{ \Omega}_{\mathfrak{X}}}  \mathcal{M} f d{\boldsymbol{ \mathfrak{X}}}
  + \nabv_{\x} \dotv (\rho D_v \nabv_{\x} q),
\end{equation}

\begin{equation}\label{eq:Vel_dim}
\begin{split}
 \rho \left(\frac{\partial \uu   }{\partial t}  
+  (\uu \dotv \nabv_{\x}) \uu \right) 
+  \nabv_{\x} p   
= &\nabv_{\x}\dotv (\mu (\nabv_{\x} \uu + \nabv_{\x}\uu^T -
\frac{2}{3}\nabv_{\x}\dotv \uu {\boldsymbol I})) \\
 & +   \rho \g \\
&  
-
\int_{\boldsymbol{ \Omega}_{\mathfrak{X}}} {\boldsymbol F}
 f  d\boldsymbol {\mathfrak{X}},
\end{split}
\end{equation}

\begin{equation}\label{eq-Tair3b} 
\begin{split}
c_\textrm{v} \rho \left( \frac{\partial T_a}{\partial t}  +  \uu \dotv\nabv_{\x} T_a \right)
  &   =   \nabv_{\x}\dotv (\kappa \nabv_{\x} T_a)  +  \taut :  \nabv_{\x} \uu  - p_a \nabv_{\x}\cdot \uu + \mathcal{D}\\
& \quad  - \int_{\boldsymbol{ \Omega}_{\mathfrak{X}}} m c_{ps}   \Trate  f  d\boldsymbol {\mathfrak{X}}\\
& \quad  + \int_{\boldsymbol{ \Omega}_{\mathfrak{X}}} L_v  \mathcal{M}  f  d\boldsymbol {\mathfrak{X}}
\end{split}
\end{equation}

We note that for all the properties of the air that we are interested in (density, humidity, velocity,
temperature), the feedback terms are integrals of the rate of
change of these quantities, for the drops, weighed by the number (concentration) distribution of the spray drops. The time rate of change of the drop velocity, mass, and temperature, are known from the microphysical equations \eqref{eq:Stokes_flow2}, \eqref{eq:DropRadius}, and \eqref{eq:DropTemp}, although they are given in a Lagrangian frame of reference (see below). Equally importantly, the distribution function $f(t,\x,\vv,r,m,T)$ now needs to be estimated.

Because spray is thought to have substantial impacts on short term weather events such as tropical storms, the airborne concentrations of drops and marine aerosols have long been of interest and several data sets have been acquired over the past decades. 
These data generally report steady and homogeneous, radius-dependent, drop concentration functions, noted $N(r)$, which represent the
number of drops per unit volume of air $d\x$ and per radius increment
$dr$.
To illustrate the model presented in this paper, we choose a somewhat reduced and idealized case in which we  consider an initial drop size distribution which is instantly introduced in the airflow at $t=0$. Here, we assume that this population of drops contains newly formed, motionless, sea water drops. In other words the initial drop distribution is:
\begin{equation}
  f(t=0,\vv,r,m,T) =f_0(\vv,r,m,T)=\delta_{\vv-0}N(r)\delta_{m-\msw(r)}\delta_{T-T_{sw}}, 
\end{equation}
with $\msw(r)=\frac{4}{3}\pi r^3\rho_{sw}$ the mass of a drop of sea water.
For the purpose of this discussion, we estimate the spray concentration $N(r)$ from the source distribution $S(r)$ using  $N(r)= \frac{S(r)}{v_{o}(r)}$, and use the source distribution outlined in \citet{Andreas:2010}, which merges the functions proposed by \citet{Fairall:1994} and \citet{Monahan:1986_dist}. 

In this example, we pick $U_{10}=15$ m s$^{-1}$ and limit our analysis to spray drops with radii from $r=1$  $\mu$m  to  $r=1$ mm. Furthermore, two-way coupling is not accounted for and the atmospheric conditions imposed on the drops are kept fixed. In other words, we consider that the drop evolve in steady atmospheric conditions and the forcing imposed on them is fixed, but we do estimate the change that the drop would otherwise induce on the atmosphere. Thus, drop velocity, mass, and temperature evolve according to the Lagrangian equations \eqref{eq:Stokes_flow2}, \eqref{eq:DropRadius}, and \eqref{eq:DropTemp}, and the drop distribution evolves following equation~\eqref{eq:WB}.
For the latter, our approach, which we detail below, is based on a semi-analytic method which uses the distribution $N(r)$ and the corresponding solution to Williams-Boltzmann equation.

Following equations~\eqref{eq-consmass_dim} to \eqref{eq-Tair3b}, at time $t$, if we ignore the effect of molecular diffusion of vapour in dry air, the spray-induced effects for mass, humidity, velocity, latent, and sensible heat, are given by:
\begin{align}
    & \MS(t)     = -\int_0^{t}
      \int_{ \Omega_{\mathfrak{X}}}  {\cal
      M}(\mathfrak{X})f(\ti,\mathfrak{X})  d{\mathfrak{X}} d\ti, \qquad
      \HS(t ) = \frac{1-q}{\rho} \MS(t), \\
    & \US(t) = -\frac{1}{\rho}\int_0^{t}
      \int_{ \Omega_{\mathfrak{X}}}  {\boldsymbol F}(\mathfrak{X})f(\ti,\mathfrak{X})  d{\mathfrak{X}}d\ti,\\
    &  \TSL(t ) = -\frac{L_v}{c_\textrm{v}\rho} \MS(t), \qquad
      \TSS(t) = -\frac{1}{c_\textrm{v}\rho}\int_0^{t}
      \int_{ \Omega_{\mathfrak{X}}}  m c_{ps}  \Trate
      (\mathfrak{X})f(\ti,\mathfrak{X})  d{\mathfrak{X}}d\ti,
     \label{eq-TSSL} 
      \end{align}
 where $f(\ti,\mathfrak{X})$ is the spatially homogeneous solution  to equation~\eqref{eq:WB} in which $\ti$ is a dummy time variable used for the integration and with $f_0$ as initial condition (see above). All feedback quantities involve integrals of the form: 
\begin{equation}\label{eq-intphi} 
   \int_{ \Omega_{\mathfrak{X}}} \phi(\mathfrak{X})f(t,\mathfrak{X})  d{\mathfrak{X}},
    \end{equation}
which can be calculated as follows. We first consolidate the Lagrangian equations~\eqref{eq:Stokes_flow2},
\eqref{eq:DropRadius},~\eqref{eq:DropTemp}, \eqref{eq-M} as a differential system,
\begin{equation}
  \frac{d\XX(t)}{dt} = \VV(\XX(t)), \qquad \XX(0) = {\mathfrak{X}}_0,
\end{equation}
where $\XX(t) = (\vv(t),r(t),m(t),T(t))$.  Using a change in variable, which for all times $t$ transforms ${\mathfrak{X}}_0=(\x_0,\vv_0,r_0,m_0,T_0)$ to ${\mathfrak{X}} = \XX(t)$, a known result from transport theory then states that:
\begin{equation}  \label{eq-transport}
 \int_{ \Omega_{\mathfrak{X}}} \phi(\mathfrak{X})f(t,\mathfrak{X})  d{\mathfrak{X}} =
 \int_{ \Omega_{\mathfrak{X}_0}} \phi(\XX(t))f_0({\mathfrak{X}_0})  d{\mathfrak{X}}_0.
  \end{equation}
The formula above then lets us solve the feedback equations above, as long as we can solve for the Lagrangian equations~\eqref{eq:Stokes_flow2},
\eqref{eq:DropRadius},~\eqref{eq:DropTemp}, \eqref{eq-M}. Because the Lagrangian equations for the spray are unnecessarily
complicated for this simple example, we use the reduced relaxation model given
in \cite{Andreas:1989}. It also has the added benefit of simple analytical solutions. 

Therefore, we approximate equations \eqref{eq:DropRadius} and ~\eqref{eq:DropTemp} with:
\begin{equation}
  \frac{dr(t)}{dt}  = \frac{r_{eq}(r_0)-r(t)}{\tau_R(r_0)},
\end{equation}
and 
\begin{equation}
  \frac{dT(t)}{dt}  = \frac{T_{ev}(r_0)-T(t)}{\tau_T(r_0)},
\end{equation}
 where $\tauR$ and $\tau_T$ are shown in figure~\ref{fig:Figure1}, $r_{eq}(r_0) = \alpha r_0$ is the radius at equilibrium ($\alpha$ depends only on ambient atmospheric conditions), and $T_{ev}$ the so-called evaporating temperature. The use of this simplified relaxation model is confirmed by examining the results in figure \ref{fig:microphys}, which shows that the temporal evolution (and thus the time rate of change) of a drop's temperature, velocity, and radius, can indeed be approximated by an exponential. Here, we consider air with a 
wind speed of $U_{10}=15$ m s$^{-1}$, a temperature of $T_a=18^oC$ and a relative humidity of $75\%$, which leads to $\alpha \approx 0.46$. The spray Lagrangian equations then yield
\begin{align}
&  u(t) = 
U_{10}-U_{10}e^{-\int_0^t\frac{ds}{\tau_D(r(s))}},\\
&   r(t) = \left(\alpha + (1-\alpha)e^{-t/\tau_R(r_0)}\right)r_0,\\
&   m(t) = m_0 + \frac{4}{3}\pi \Bigl( \left(\alpha + (1-\alpha)e^{-t/\tau_R(r_0)} \right)^3-1\Bigr)r_0^3\rho_w,\\
&   T(t) = T_{ev}(r_0) + (T_0-T_{ev}(r_0))e^{-t/\tau_T(r_0)},
\end{align}
where, for simplicity, we only consider the horizontal component of the velocity $u(t)$. The corresponding Lagrangian rates of changes are
\begin{align}
\label{eq-relax_U}
  & F(t) = \frac{m(t)}{\tau_D(r(t))}(U_{10}-u(t)) ,  \\ \label{eq-relax_R}
  & {\cal R}(t) = \frac{\alpha r_0 - r(t)}{\tau_R(r_0)}, \\ 
  \label{eq-relax_M}
  & {\cal M}(t) = 4\pi r(t)^2 \rho_{w} \mathcal{R}(t) , \\
  \label{eq-relax_T}
  & {\cal T}(t) = \frac{T_{ev}(r_0)-T(t)}{\tau_T(r_0)}. 
\end{align}
where equation~(\ref{eq-transport}) naturally leads to a relatively simple estimate of the airside variations in density, humidity, temperature, and speed, which are induced by the introduction in the airflow of a population of sea-water spray drops at time $t=0$.

\subsubsection{Estimation of the feedback on the humidity of the air}

From equation~\eqref{eq-consmass_dim}, the water mass (per unit volume. i.e. the density) transferred to the air by the evaporating spray is
\begin{equation}
\MS(t)     = -\int_0^{t}
      \int_{ \Omega_{\mathfrak{X}_0}}  {\cal
      M}(\ti)f_0(\mathfrak{X}_0)  d
    {\mathfrak{X}}_0d\ti,
\end{equation}
in which  ${\cal
      M}(\ti)$ is given by equation \eqref{eq-relax_M}  above. Using the known value for $f_0$ and integrating in time  yields
    \begin{equation}  \label{eq-Ms}
\MS(t)     = \int_0^{+\infty} \frac{4}{3}\pi r_0^3\rho_w
\left(1-\alpha^3-\phi \left(\frac{t}{\tau_R(r_0)}\right)\right)N(r_0)\, dr_0,
\end{equation}
where $\phi(\zeta) = (1-\alpha)(3\alpha^2e^{-\zeta}+3\alpha(1-\alpha)e^{-2\zeta}
+ (1-\alpha)^2e^{-3\zeta})$.

Note that for large times, $\MS$ tends toward
\begin{equation}
  (1-\alpha^3)\int_0^{+\infty} \frac{4}{3}\pi r_0^3\rho_w
N(r_0)\, dr_0,
  \end{equation}
which simply represent the initial pure water mass, less the maximum water mass that can potentially evaporate, $\int_0^{+\infty} \frac{4}{3}\pi (\alpha r_0)^3\rho_w
N(r_0)\, dr_0$. In fact, this limit readily gives a quick upper limit estimate of the potential total water vapour that can be transferred to the atmosphere.

\begin{figure}
  \centering
  \includegraphics[width=0.65\textwidth]{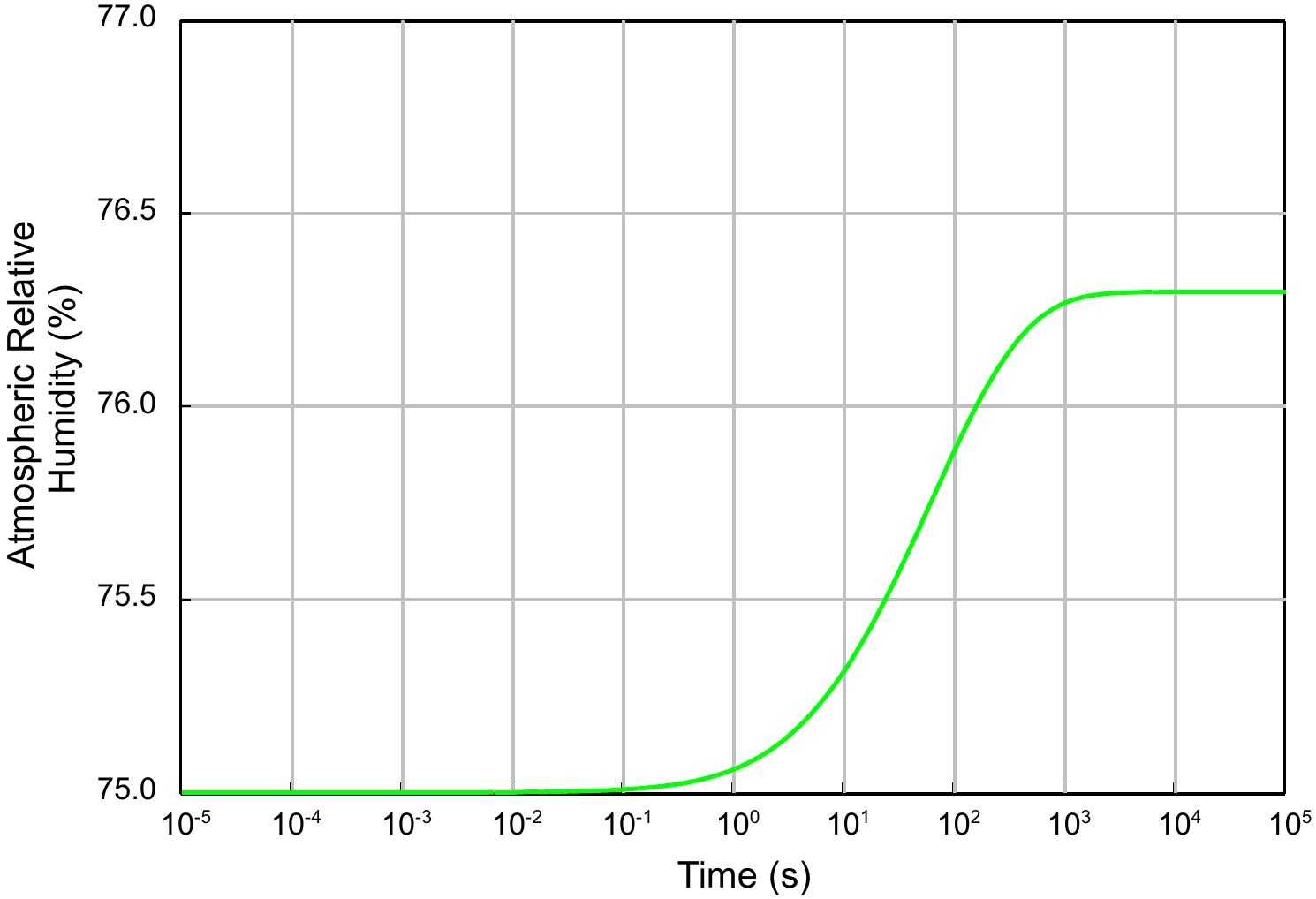}
  \caption{Relative
  humidity due to the spray drops with a size distribution $N(r)$. The distribution is assumed to be introduced at $t=0$ in air with a steady wind speed of $U_{10}=15$ m s$^{-1}$, a temperature of $T_a=18^oC$, and a relative humidity of $75\%$. The drops all have an initial temperature of $T_p=20^oC$.}
  \label{fig:Humidity}
\end{figure}

In figure~\ref{fig:Humidity}, we plot the corresponding relative
humidity increase as a function of time, assuming that the
spray drop distribution is introduced in the airflow at
$t=0$. Similarly to the flow conditions of figure~\ref{fig:microphys},
the air has a steady wind speed of $U_{10}=15$ m s$^{-1}$, a
temperature of $T_a=18^oC$, and a relative humidity of $75\%$. The
drops all have an initial temperature of $T_p=20^oC$. Within $O(10^3)$
s, the drop distribution has transferred the bulk of its water vapour
to the atmosphere. During that time, the air relative humidity has
increased from $75\%$ to approximately $76.3\%$. We note here that
figure~\ref{fig:Humidity} shows the aggregated results for the whole
size distribution $N(r)$. In figure \ref{fig:Figure1}, the relaxation
time $\taum$ for this distribution ranges from $O(10^{-2})$ s to
$O(10^4)$ s and we therefore expect the small drops to evaporate  much
faster than the large drops. But the time over which atmospheric
humidity is increased which is depicted in figure~\ref{fig:Humidity},
is due to the whole  size dependent distribution $N(r)$, along with the
corresponding size dependent relaxation time $\taum(r)$. In fact, it
appears that the distribution $N(r)$ does not possess a sufficient
amount of large drops of sizes $O(1)$ mm to drive the relative
humidity increase over long times, and the relative humidity
stabilizes at times shorter than $O(10^4)$ s, time that would be
required for the largest drops to evaporate. In other words, the
larger drops in the distribution do not  contribute much to the change
in humidity; if they did, the atmospheric humidity would still be
changing until times of order $O(10^4)$ s.

This is somewhat fortuitous
because the model above also assumes that all the drops of the
distribution remain suspended in the air. Furthermore, two-way coupling was not accounted for in this idealized example and the atmospheric conditions are considered steady. In reality, ambient relative humidity changes as the spray evaporates. Of course, the fully unsteady solutions are obtained by solving equation \eqref{eq:WB} which gives the spray size distribution, concurrently with equations  \eqref{eq-consmass_dim} or \eqref{eq-q_dim}, \eqref{eq:Vel_dim} and  \eqref{eq-Tair3b} which give the atmospheric condition including the (instantaneous) spray effects.
In that sense, the result above is a slight overestimate of the the feedback effects from drop distribution on the atmospheric relative humidity. 


\subsubsection{Estimation of the feedback on the air velocity}

Similarly, from equation ~\eqref{eq-transport}, at time $t$, the variation in wind speed due to the spray distribution is
\begin{equation}
  \begin{split}
\US(t)     & = -\frac{1}{\rho}\int_0^{t}
      \int_{ \Omega_{\mathfrak{X}_0}}  F(\ti)f_0(\mathfrak{X}_0)  d
      {\mathfrak{X}}_0 d \ti\\
      & = -\frac{1}{\rho}\int_0^{t}
      \int_{ \Omega_{\mathfrak{X}_0}}   \frac{\frac{4}{3}\pi
        r(\ti)^3\rho_w + \msw(r_0)}{\tau_D(r(\ti))}
      U_{10}  e^{-\int_0^{\ti}\frac{ds}{\tau_D(r(\ti))}} N(r_0)
      \, dr_0 d\ti.   
  \end{split}
\end{equation}
Neglecting high Reynolds number corrections on the drag force, thus assuming that
$\tau_D(r) = \tau_{p}(r) = \frac{2r^2\rho_{sw}}{9\mu}$ (see section~\ref{subsec:motion}), and noting that
$\frac{\tau_D}{\tau_R}\approx O(10^{-3})\ll 1$ (see figure~\ref{fig:Figure1}), lead to a simpler analytical estimate
\begin{equation}
  \US(t)= -\frac{1}{\rho}U_{10}\alpha^2\left(1+\frac{\rho_w}{\rho_{sw}}\right)
      \int_{ 0}^{+\infty}  \msw(r_0)\left(1-e^{-\frac{t}{\alpha^2\tau_D(r_0)}}\right)  \, N(r_0) dr_0.
    \end{equation}

Note that for large times, $\US$ tends toward:
    \begin{equation}
       -\frac{1}{\rho}U_{10}\alpha^2\left(1+\frac{\rho_w}{\rho_{sw}}\right)
      \int_{ 0}^{+\infty}  \msw(r_0)N(r_0) \, dr_0,
    \end{equation}
    where $\int_{
      0}^{+\infty}  \msw(r_0)N(r_0) \, dr_0$ is sea spray mass fraction.
Again, this gives a quick way to estimate the potential dynamic effect of a given spray concentration $N(r_0)$.

\begin{figure}
  \centering
  \includegraphics[width=0.65\textwidth]{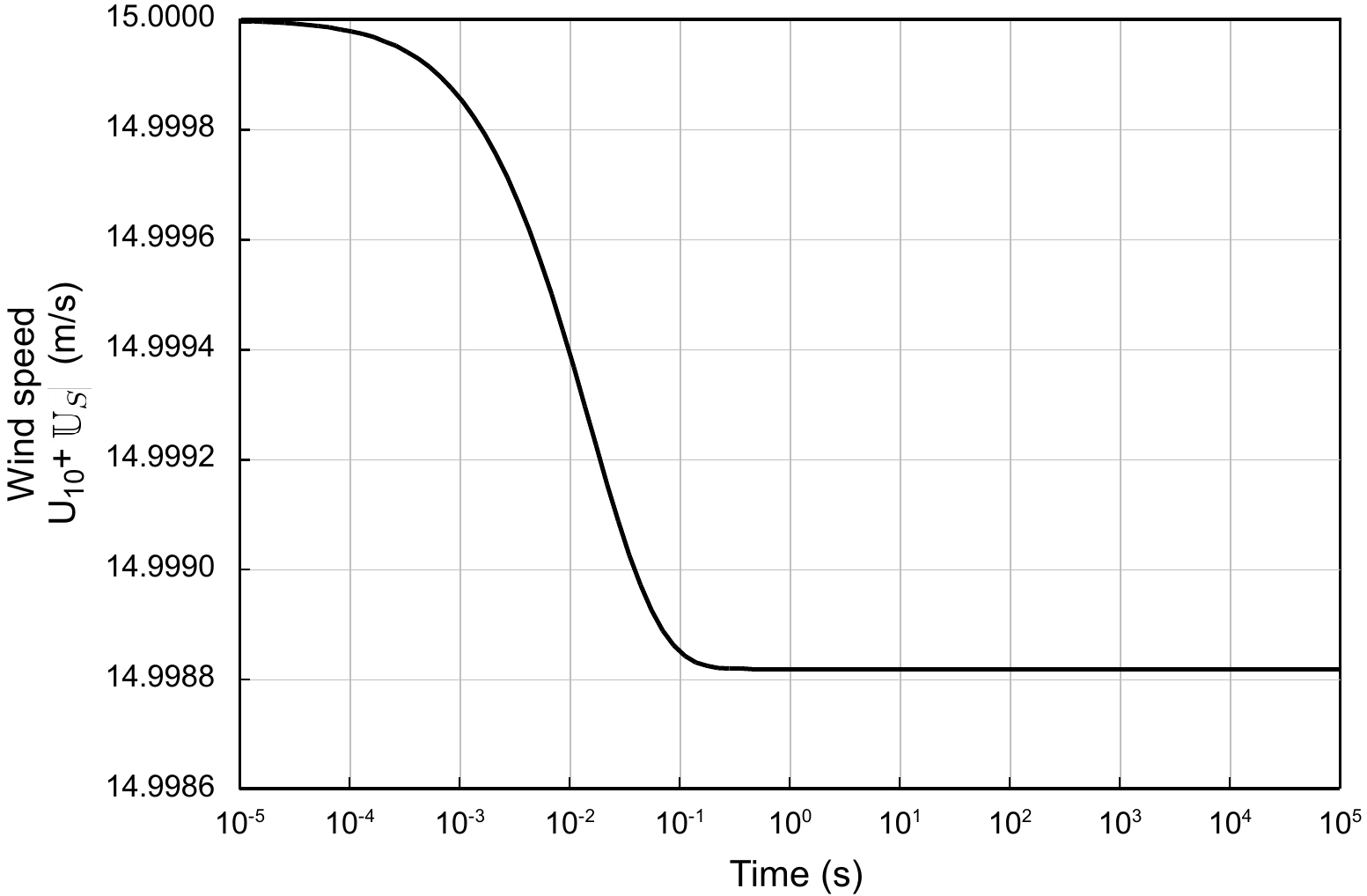}
  \caption{Decrease of the air velocity due to the drag force
    exerted by the drops given by the distribution $N(r)$. Initial ambient conditions are the same as those shown in figure \ref{fig:Humidity}}
\label{fig:velocity_feedback}
\end{figure}

 Figure~\ref{fig:velocity_feedback} shows the variation in wind speed due to the spray distribution. For the exchange of momentum, figure \ref{fig:Figure1} indicates that the relaxation time is sufficiently short for most of the drop sizes to fully exchange momentum with the atmosphere. In fact, figure \ref{fig:velocity_feedback} shows that the whole distribution has completed its momentum exchange in less that $O(1)$ s. Yet, the momentum extracted from the wind to accelerate the spray drops (which are assumed to start at rest) is negligible, at least in the example taken here. This simply indicates that the spray mass loading given by $N(r_0)$ at this wind speed is too small to have a dynamic effect on the carrier fluid. Larger mass loading evidently can have a substantial impact on the spray-atmosphere momentum exchange (see for example the recent work of \citet{Richter:2013, RichterPoF:2013, Richter:2015, Richter:2016}).

\subsubsection{Estimation of the feedback on the air temperature}

From equation~\eqref{eq-Tair3b}, the rate of change of the air temperature due to the evaporation of sea spray is the sum of the two terms corresponding to the spray latent and sensible heat transfers. 
Together, equations~(\ref{eq-TSSL}) and~(\ref{eq-Ms}) give the spray induced latent heat $\TSL(t )$. Likewise, equations~(\ref{eq-TSSL}) and~\eqref{eq-transport} give the component corresponding to the spray sensible heat
\begin{equation}
  \begin{split}
\TSS(t) & = -\frac{1}{c_\textrm{v}\rho} \int_0^{t}
      \int_{ \Omega_{\mathfrak{X}_0}}  m(\ti) c_{ps}{\cal
      T}(\ti)f_0(\mathfrak{X}_0)  d
    {\mathfrak{X}}_0d\ti \\
    & = \frac{1}{c_\textrm{v}\rho}
    \int_{0}^{+\infty} \left(\msw(r_0)(1-e^{-t/\tau_T(r_0)}) \right.\\
      & \qquad \qquad \qquad 
      \left. + \frac{4}{3}\pi r_0^3\rho_w\psi\left(\frac{t}{\tau_T(r_0)},\frac{\tau_T(r_0)}{\tau_R(r_0)}\right)
    \right) c_{ps} (T_{sw}-T_{ev}(r_0))N(r_0)\, dr_0,
  \end{split}
\end{equation}
with
\begin{equation}
  \begin{split}
    \psi(x,y) = & (\alpha^3-1)(1-e^{-x})
    + 3\alpha^2(\alpha-1)\frac{1}{1+y}(1-e^{-x(1+y)})  \\
    & + 3 \alpha(\alpha-1)^2\frac{1}{1+2y}(1-e^{-x(1+2y)})
    + (1-\alpha)^3\frac{1}{1+3y}(1-e^{-x(1+3y)}).
  \end{split}
  \end{equation}

 Noting that $\frac{\tau_T}{\tau_R}\approx O(10^{-3})\ll 1$ (see
  figure~\ref{fig:Figure1}), the expression above can be substantially simplified:
  \begin{equation}
    \begin{split}
\TSS(t) = \frac{1}{c_\textrm{v}\rho}
    \int_{0}^{+\infty} & \left(\msw(r_0) - 6\alpha^2(1-\alpha)\frac{4}{3}\pi
      r_0^3 \rho_w    \right) (1-e^{-t/\tau_T(r_0)})\\
    & c_{ps} (T_{sw}-T_{ev}(r_0))N(r_0)\, dr_0.
  \end{split}
\end{equation}

\begin{figure}
  \centering
  \includegraphics[width=0.75\textwidth]{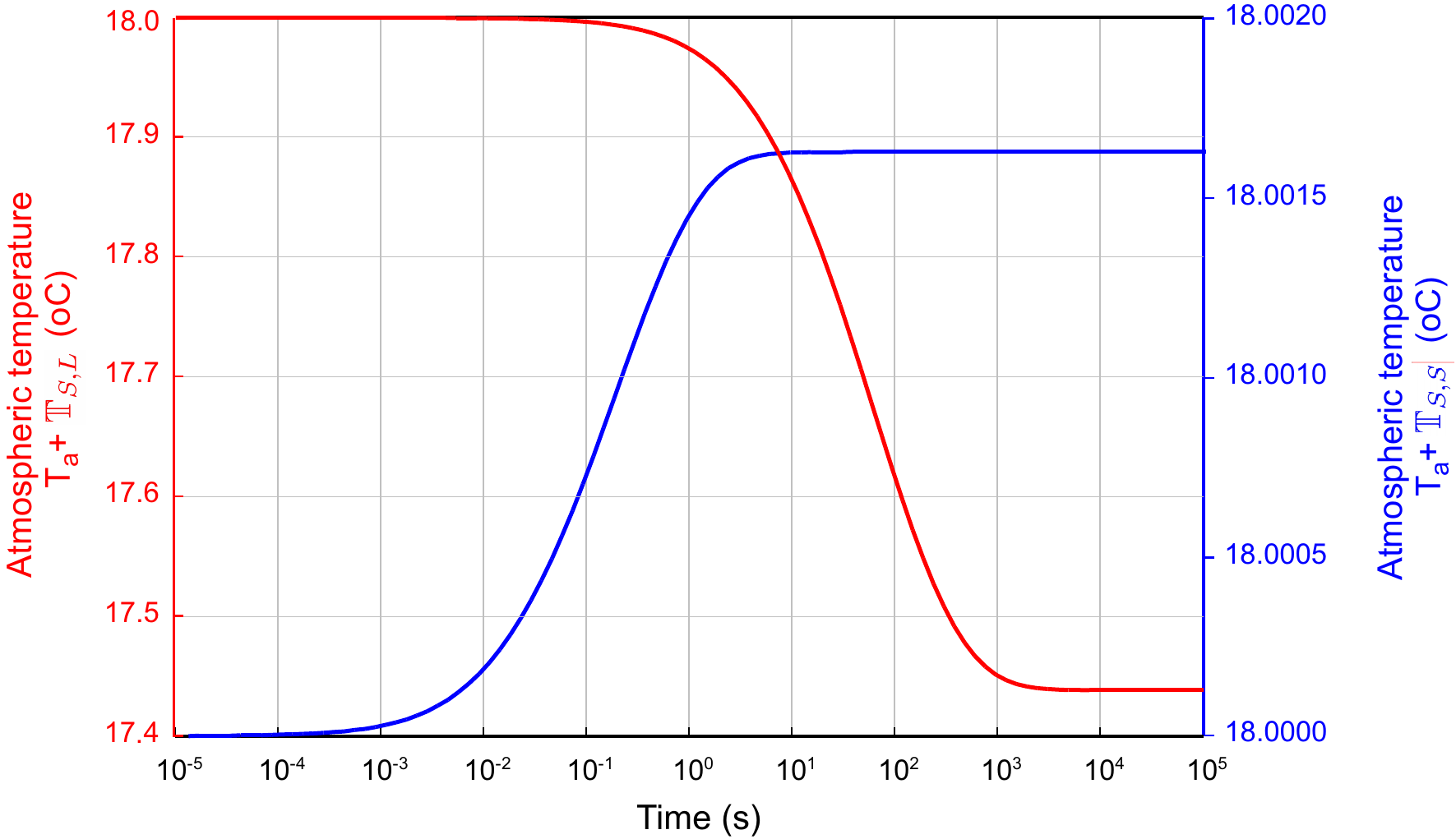}
  \caption{Variation of the air temperature due to the sensible (blue) and latent (red) heat flux from the distribution of sea spray drops given by $N(r)$. Initial ambient conditions are the same as those shown in figure \ref{fig:Humidity} }
\label{fig:temperature_feedback}
\end{figure}

 Figure~\ref{fig:temperature_feedback} shows the resulting variation in the air temperature induced respectively by the latent (red curve) and sensible (blue curve) heat exchanges from the spray. Again, just as with the spray effects on the humidity, the relaxation time for the latent heat flux is such that the spray contribution may be slightly overestimated when accounting for the entire size distribution including few large drops. The temperature change induced by the spray sensible flux is likely to be accurately accounted for, but the results show that it is a negligible part of the total spray-induced air temperature change. Indeed, figure~\ref{fig:temperature_feedback} shows that the bulk to the spray effects on the air temperature comes from the latent heat exchange from the drop distribution. As such, just like the effects on the relative humidity, it also takes $O(10^3)$ s to fully establish. In the example taken here, the atmospheric temperature is decreased by approximately $0.5^oC$.

 \subsection{Perspectives}

Lastly, we offer here a discussion of possible future work and potential avenues for improvements to the preliminary model presented above.


\subsubsection{Incompressible approximation}

While the approach presented here can be applied to a range of particle-laden flows, in the case of sea spray, the air can generally be  considered as nearly incompressible. Indeed, the Mach number ranges from 0.01 for a wind speed of 5 m \;s$^{-1}$ to approximately 0.2 in extreme winds of 80 m \;s$^{-1}$. In this case, it is customary to replace the Navier-Stokes equations by the corresponding incompressible equations. Formally, these equations are obtained by assuming a uniform and steady density for the air, in which case the conservation equation of the mass is replaced by $\nabv _{\x}\dotv  \uu = 0$; this new relationship is then included in the momentum equation. Furthermore, if the variations of temperature are sufficiently low (of the order of a few percent in Kelvin, or less than a dozen degrees Celsius, which is generally the case for sea spray), one can take into account the natural convection by using the well-known Boussinesq approximation in which the gravity term $\rho
\g$ is replaced with $\frac{T-\Tchar}{\Tchar}\rhochar g$. In our case, terms accounting for feedback effects from the spray complicate the traditional Boussinesq approximation. Therefore, for readers interested in pursuing the analysis beyond what is presented here, we recommend proceeding with the so called ``low Mach'' asymptotic analysis~\citep{book_feireisl}  where equations are written as a series expansion using the Mach number as the small parameter.



\subsubsection{Turbulent flows}

The airflow over the wavy ocean surface is turbulent and the 
 fluctuations in the air velocity thereby induce  fluctuations in the
velocity of the drops. When the Navier-Stokes equations for the airflow are solved with a sufficient resolution, the present model accounts for these fluctuations. However, solving Navier-Stokes equations with domains sufficiently large to cover spray transport generally requires numerical resolutions which are not sufficient to capture the
turbulence of the air, which is then typically modeled using LES or RANS approaches. In this case, the diffusion induced by the turbulence is parametrized without specifically describing the fluctuations of the
velocity field. Equivalently, our model can be modified to account for the turbulent fluctuations of the air and drop velocities. In the Lagrangian description, fluctuations for the drop velocities can be
modeled by a turbulence-induced forcing thus leading to a stochastic differential equation (analogous to a Langevin's equation) for $\vp $.
In the Eulerian framework, a corresponding (and still deterministic)
Williams-Boltzmann equation can be obtained
through the Feynman-Kac representation theorem (see for example ~\citet{Domelevo:2007}).



\subsubsection{Numerical simulations}

The Eulerian model presented here to study spray and spray effects on the atmosphere is consistent with the typical Eulerian approach for solving large-scale atmospheric flows. From a numerical point of view, this simply means that similar numerical methods can be used for the spray, the carrier atmospheric flow, and the feedback/coupling terms. Meanwhile, using a Lagrangian model to describe the spray and an Eulerian model
for the carrier fluid is an approach that is frequently used \citep[e.g.][]{Edson:1994, Edson:1996, Mueller:2014a, Mueller:2014b, Richter:2013, RichterPoF:2013, HelgansIJMF:2016, RichterPoF:2014, Peng:2019, Richter:2019}, but which presents a number of challenges such as
numerical stability problems and a computing cost that can quickly become
prohibitive especially for unsteady flows.

Practically, an Eulerian numerical simulation will not directly solve the complete Williams-Boltzmann (equation~\ref{eq:WB}) because
the size of the phase space is generally far too large (at least
9 variables in a three-dimensional physical space). Thus, so called
``fluid models'' obtained from the kinetic equation are generally favored. These
models can be obtained from conservation equations for
 moments of the distribution function  $f(t,\x,\vv,r,m,T)$,  with the addition of an appropriate closure hypothesis.
This approach is used successfully in combustion~\citep{DFV_2008,livre-Cox}, and several types of models exist. In
particular, the ``Quadrature-Based Moment Method'' approach seems
quite suitable for the spray problem presented here. For example, we briefly describe below the single-node closure in which the distribution $ f(t,\x,\vv,r,m,T) $ is reduced to
$n(t,\x,r,m)\delta_{\vv-\UU(t,\x,r,m)}\delta_{T-\Theta(t,\x,r,m)}$. This simply means that at any point in space, drops of radii $ r $ and mass $ m $  all have the same  speed $ \UU $ and
the same temperature $\Theta $, while these parameters can change in time and space. 
With these simplifications,
it is straightforward to obtain three evolution equations for $ n $, $ \UU $
and $ \Theta $. Indeed, integrating~(\ref{eq:WB}) multiplied by $ 1 $, $ \vv $, and $ T $, yields:
\begin{align}
&    \frac{\partial n}{\partial t} + \nabv_{\x}\dotv (n \UU)
    +\frac{\partial }{\partial r}(\bar{\mathcal{R}}n)
    +\frac{\partial }{\partial m}(\bar{\mathcal{M}}n) =0, \\
&      \frac{\partial n\UU}{\partial t} + \nabv_{\x}\dotv (n \UU\otimes \UU)
    +\frac{\partial }{\partial r}(\bar{\mathcal{R}}n\UU)
    +\frac{\partial }{\partial m}(\bar{\mathcal{M}}n\UU) 
     = n\g + \frac{1}{\tauI}n(\uu-\UU), \\
&   \frac{\partial n\Theta}{\partial t} + \nabv_{\x} \dotv (n \UU\Theta)
    +\frac{\partial }{\partial r}(\bar{\mathcal{R}}n\Theta)
    +\frac{\partial }{\partial m}(\bar{\mathcal{M}}n\Theta) 
     = \bar{\Trate}n, 
\end{align}
where $\bar{\mathcal{R}}$, $\bar{\mathcal{M}}$, and  $\bar{\Trate}$ are the rate of variations of the drops in which the kinetic variables $\vv$ and $T$ are replaced by the values $\UU$ and $\Theta$ of the closure. This system can be used instead of the complete Williams-Boltzmann equation, which is no longer necessary. 

The feedback terms are also easily obtained from $ n $, $ \UU $, and $ \Theta $. Thus, the reduced model
consists of semi-macroscopic equations with $ n $, $ \UU $, and $ \Theta $ above and
the equations for the air flow.  It is then possible to
further reduce the model by eliminating the variables $ r $
and $ m $. To that end, two types of approaches exist. The first is
is based on a direct integration and leads to the loss of the polydispersed nature of the problem; this obviously seems inappropriate in the case of sea spray. The second is to discretize the variables $ r $
and $ m $, as in the ``sampling'' and ``sectional'' or ``multi-fluid'' approaches; this seems better suited for the problem at hand (see~\cite{livre-Cox}). 

For realistic applications, this one node closure is not sufficient, and several higher order closure have been recently developped to obtain more accurate models, see~\cite{FLV_2018} and the references therein.

\section{Conclusions}
 
In this paper, we have presented a novel model to estimate sea spray feedback effects on the atmospheric flow. Evidently, the spray effects depend on the exchange rates of mass, momentum, and heat, from the population of drops that make up sea spray. The spray mediated exchanges are therefore generally obtained from the equation of motion and the thermodynamics of individual drops as they are transported in the atmospheric boundary layer. These are naturally given from a Lagrangian point of view. In fact, Lagrangian drop transport models are valuable tool to estimate spray residence times in the air.
However, the atmospheric flow is better described in a conventional Eulerian frame of reference with the usual mass, momentum, and energy conservation equations, i.e. the standard Navier-Stokes set.
Evaluating spray effects on the atmosphere thus generally involves coupling the Lagrangian spray transport with the Eulerian atmospheric flow. At best, this is largely impractical and computational costly.
In this paper, we present a model which utilizes the kinetic theory framework and thus reconciles these two approaches. This kinetic approach affords substantial advantages as it leads to a coupled Eulerian spray-atmosphere model which can be solved to estimate spray feedback effects.

We further performed a dimensional analysis and isolated the spray induced dominant terms in the equations for the atmospheric flow.  Subsequently, we presented, for illustration purpose, a single estimate of the spray feedback effects in which we have used a commonly accepted sea spray drop size distribution forcing function. A complete investigation of the spray effects over a wide range of atmospheric humidity, temperature, and wind speed is beyond the scope of this paper.
Nevertheless, the spray induced feedback on the atmosphere which results from this simple estimate, and for these particular conditions, shows that the spray latent flux, but not the sensible flux, drives an atmospheric temperature change. The spray evaporation also has a measurable effect on the relative humidity in the airflow. The spray effects on the momentum of the airflow are negligible in the condition studied. We expect spray momentum effect to occur only at substantial spray mass loading.

While the model developed here is robust, the largest source of error and uncertainty lies with the difficulty of defining both a reliable spray size distribution and accurate spray residence times in the airflow. Clearly, more data are needed on both accounts. Experiments are needed in order to obtain, over a wide range of environmental conditions, measured spray size distributions, or spray generation functions which are generally needed to assess air-sea fluxes.\\
\\

\section*{Acknowledgements}
This work was supported in part on grant OCE-1829660 from the US National Science Foundation.$\vchars$

\nomenclature{$\x$}{Position vector (Eulerian)}%
\nomenclature{$\xp$}{Drop position vector (Lagrangian)}%
\nomenclature{$\vv$}{Drop velocity (Eulerian)}
\nomenclature{$\vp$}{Drop velocity (Lagrangian)}
\nomenclature{$\uu$}{Fluid velocity (Eulerian) }%
\nomenclature{$\up$}{Fluid velocity (Lagrangian)}%

\nomenclature{$\Uv$}{Molecular velocity of water vapour }%
\nomenclature{$\Ua$}{Molecular velocity of dry air }%

\nomenclature{$m_p$}{Drop mass}%
\nomenclature{$r_p$}{Drop radius}%
\nomenclature{$r_{p0}$}{Drop radius at formation}%
\nomenclature{$m_{p0}$}{Drop mass at formation}%
\nomenclature{$T_p$}{Drop temperature}%
\nomenclature{$E_p$}{Drop energy}%
\nomenclature{$e_p$}{Drop specific internal energy}%

\nomenclature{$\rho_p$}{Drop density}%
\nomenclature{$\boldsymbol {\mathcal{F}}$}{Forces on a drop}%
\nomenclature{$\boldsymbol F$}{Friction forces on a drop}%
\nomenclature{$C_D$}{Drag coefficient on a drop}%
\nomenclature{$A$}{Drop Cross-sectional area}%
\nomenclature{$\Xi$}{Drag coefficient correction for high drop Reynolds number}%
\nomenclature{$\nu$}{Air kinematic viscosity}%
\nomenclature{$\mu$}{Air dynamic viscosity}%

\nomenclature{$f_w$}{Correction factor (ventilation coefficient) for water vapour diffusivity}%
\nomenclature{$Q_{RH}$}{Fractional relative humidity}%
\nomenclature{$T_a$}{Air temperature}%
\nomenclature{$R$}{Universal gas constant}
\nomenclature{$M_w$}{Molecular weight for water}
\nomenclature{$M_s$}{Molecular weight for NaCl salt}
\nomenclature{$D_v^*$}{Water vapour diffusivity corrected for non-continuum effects}
\nomenclature{$L_v$}{Latent heat of vaporisation}
\nomenclature{$p_v^{sat}$}{Saturation vapour pressure}
\nomenclature{$\Gamma_p$}{Surface tension}
\nomenclature{$\Phi_s$}{Osmotic coefficient}
\nomenclature{$m_s$}{Mass of salt in a spray drop}
\nomenclature{$e_{sw}$}{Specific internal energy of sea water}%

\nomenclature{$c_{ps}$}{Specific heat of salty water}
\nomenclature{$f_h$}{Correction factor (ventilation coefficient) for Thermal conductivity}%
\nomenclature{$k^*_a$}{Thermal conductivity of air corrected for non-continuum effects}
\nomenclature{$q$}{Specific humidity for moist air}

\nomenclature{$k_a$}{Thermal conductivity of air}
\nomenclature{$c_{\textrm{v},a}$}{Specific heat of dry air at constant volume}
\nomenclature{$c_{\textrm{p},a}$}{Specific heat of dry air at constant pressure}
\nomenclature{$c_{\textrm{v},v}$}{Specific heat of water vapour at constant volume}
\nomenclature{$c_{\textrm{p},v}$}{Specific heat of water vapour air at constant pressure}
\nomenclature{$c_{\textrm{v}}$}{Specific heat of  air at constant volume}
\nomenclature{$c_{\textrm{p}}$}{Specific heat of  air at constant pressure}

\nomenclature{$\mathcal{M}$}{Time rate of change of the mass of a drop}
\nomenclature{$\mathcal{R}$}{Time rate of change of the radius of a drop}
\nomenclature{$\mathcal{T}$}{Time rate of change of the temperature of a drop}
\nomenclature{$\mathcal{E}$}{Time rate of change of the energy of a drop}

\nomenclature{$f$}{Drop distribution function}

\nomenclature{$\uu_s$}{Velocity of the sea surface}
\nomenclature{$v_o$}{Drop vertical deposition velocity, or drop velocity scale}
\nomenclature{$\textrm{v}_t$}{Drop vertical terminal velocity}

\nomenclature{$\boldsymbol{n}$}{Unit vector normal to the sea surface}
\nomenclature{$S$}{Spray generation function}
\nomenclature{$N$}{Spray number concentration function}

\nomenclature{$\boldsymbol{\tau}$}{Air stress tensor}
\nomenclature{$\boldsymbol{\sigma}$}{Air viscous stress tensor}
\nomenclature{$\boldsymbol{Q}$}{Atmospheric heat flux}
\nomenclature{$\mathsf{E}$}{Atmospheric energy density}
\nomenclature{$\mathsf{E}_v$}{Energy density of water vapour}
\nomenclature{$\mathsf{E}_a$}{Energy density of dry air}
\nomenclature{$p$}{Atmospheric pressure}
\nomenclature{$\g$}{Gravity}

\nomenclature{$\mathsf{e}$}{Specific internal energy of air}
\nomenclature{$\mathsf{e}_a$}{Specific internal energy of dry air}
\nomenclature{$\mathsf{e}_v$}{Specific internal energy of water vapour}
\nomenclature{$h_a$}{Specific enthalpy of dry air}
\nomenclature{$h_v$}{Specific enthalpy of water vapour}

\nomenclature{$\rhosw$}{Density of sea water}%
\nomenclature{$\rhow$}{Density of pure water}%
\nomenclature{$\rho$}{Density of moist air}%
\nomenclature{$\rho_a$}{Density of dry air}%
\nomenclature{$\rho_v$}{Water vapour density}%
\nomenclature{$p$}{Atmospheric pressure}%
\nomenclature{$p_a$}{Dry air pressure}%
\nomenclature{$p_v$}{Water vapour pressure}%
\nomenclature{$\vchars$}{Characteristic drop velocity}%
\nomenclature{$\xchar$}{Characteristic length scale}%
\nomenclature{$\uchar$}{Characteristic air velocity scale}%
\nomenclature{$\Tchar$}{Characteristic temperature scale}%
\nomenclature{$\ratiov$}{Ratio of the characteristic drop and air flow velocities}%
\nomenclature{$\taugrav$}{Characteristic drop time scale from gravity}

\nomenclature{$[.]'$}{Primes $'$ denotes dimensionless values}
\nomenclature{$[.]_o'$}{Subscript $_o$ denotes charateristic values}

\nomenclature{$\tauR$}{Relaxation time for the drop radius}
\nomenclature{$\taum$}{Relaxation time for the drop mass}
\nomenclature{$\tauT$}{Relaxation time for the drop temperature}
\nomenclature{$\tauI$}{Inertial relaxation time or drop inertial time scale}


\nomenclature{$\aI$}{Dimensionless inertial relaxation time}
\nomenclature{$\aR$}{Dimensionless relaxation time for drop radius}
\nomenclature{$\am$}{Dimensionless relaxation time for drop mass}
\nomenclature{$\aT$}{Dimensionless relaxation time for drop temperature}

\nomenclature{$\Ma$}{Mach number}
\nomenclature{$\Reyn$}{Reynolds number}
\nomenclature{$\Pe$}{Peclet number for heat}
\nomenclature{$\Pem$}{Peclet number for water vapour}
\nomenclature{$\cchar$}{Speed of sound}

\nomenclature{$\prodm$}{Exchange rate of mass between the drop distribution and the air}
\nomenclature{$\produ$}{Exchange rate of momentum  between the drop distribution and the air}
\nomenclature{$\prode$}{Exchange rate of energy  between the drop distribution and the air}

\nomenclature{$\MS$}{Variation in air density due to the spray drops}
\nomenclature{$\HS$}{Variation in air humidity due to the spray drops}
\nomenclature{$\US$}{Variation in air velocity due to the spray drops}
\nomenclature{$\TSL$}{Variation in air latent heat due to the spray drops}
\nomenclature{$\TSS$}{Variation in air sensible heat  due to the spray drops}

\printnomenclature

\newpage
\bibliographystyle{jfm}
\bibliography{references_Aug2018}

\end{document}